\makeatletter
\@ifundefined{@parse@version@dash}{%
\def\@parse@version#1{\@parse@version@0#1}%
\def\@parse@version@#1/#2/#3#4#5\@nil{%
  \@parse@version@dash#1-#2-#3#4\@nil}%
\def\@parse@version@dash#1-#2-#3#4#5\@nil{%
  \if\relax#2\relax\else#1\fi#2#3#4}%
}{}%
\makeatother

\documentclass[
 reprint,
superscriptaddress,
longbibliography,
 amsmath,amssymb,
 aps,
]{revtex4-2}

\usepackage[T1]{fontenc}
\usepackage[utf8]{inputenc}
\usepackage{graphicx}
\usepackage{dcolumn}
\usepackage{bm}
\usepackage[normalem]{ulem}
\usepackage{cancel}

\usepackage{comment}
\usepackage[sectionbib]{bibunits}
\defaultbibliographystyle{apsrev4-2}

\usepackage[colorlinks=true,linkcolor=blue,citecolor=blue,urlcolor=blue]{hyperref} 
\usepackage{booktabs}
\usepackage{array}
\usepackage{ragged2e}
\usepackage{amsmath}
\usepackage{braket}
\usepackage{placeins}
\usepackage{xcolor}
\usepackage{capt-of}

\makeatletter
\newcommand{\smcontentsline}[3]{%
  \noindent
  \hyperref[#3]{#1\quad #2}%
  \dotfill
  \pageref{#3}\par
}
\makeatother

\begin{document}

\preprint{APS/123-QED}


\title{Emergence and suppression of phonon vortices in two-dimensional crystals: Interplay of lattice symmetry, heavy impurities, and shear}

\author{Yu-Tian Zhang}
 \affiliation{Institute of Theoretical Physics, Chinese Academy of Sciences, Beijing 100190, China}
 \affiliation{School of Physical Sciences, University of Chinese Academy of Sciences, Beijing 100049, China}

\author{Deng Pan}
 \affiliation{Institute of Theoretical Physics, Chinese Academy of Sciences, Beijing 100190, China}
 \affiliation{School of Physics and Information Technology, Shaanxi Normal University, Xi’an 710119, China}

\author{Yuliang Jin}
    \email[Corresponding author: ]{jinyuliang@itp.ac.cn} 
 \affiliation{Institute of Theoretical Physics, Chinese Academy of Sciences, Beijing 100190, China}
 \affiliation{School of Physical Sciences, University of Chinese Academy of Sciences, Beijing 100049, China}
 \affiliation{Center for Theoretical Interdisciplinary Sciences, Wenzhou Institute, University of Chinese Academy of Sciences, Wenzhou, Zhejiang 325001, China}

\date{August 31, 2026}


\begin{abstract}
Phonon vortices are vortex-like displacement fields that appear in the vibrational modes of two-dimensional materials. Here, we demonstrate that these vortices arise as symmetry-adapted linear combinations of degenerate planar phonon modes, with the superposition coefficients uniquely determined by the lattice point group. This symmetry principle establishes that vortices are intrinsic standing-wave solutions in perfect crystals, requiring neither impurities nor disorder. A heavy mass impurity favors vortex modes over planar modes through stronger resonance-induced frequency softening, whereas shear deformation suppresses vortex modes by breaking rotational symmetry. The competition between these two effects gives rise to a precisely predictable strain threshold. 
The proposed framework, grounded in symmetry and energy-minimization, provides a useful basis for understanding vibrational topological defects induced by other types of impurities or defects. This study further suggests that defect and shear engineering constitutes an effective tuning strategy for controlling vibrational modes and the associated thermal and mechanical properties of crystals.
\end{abstract}

\maketitle


\begin{bibunit}[apsrev4-2]

\enlargethispage{2\baselineskip}

\begin{figure*}[t!]
\centering
\includegraphics[width=0.97\textwidth]{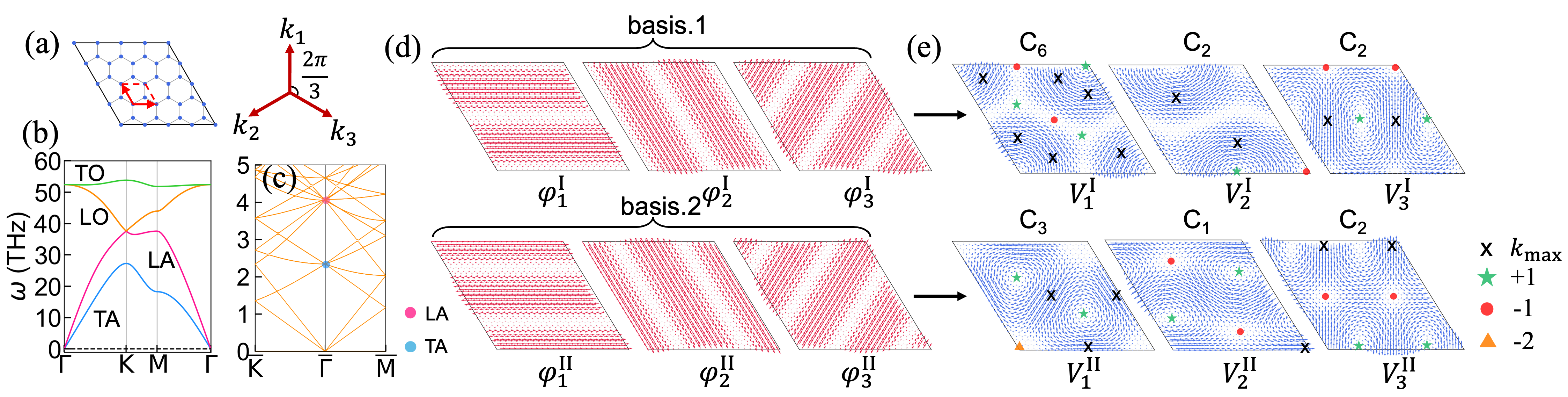}
\caption{{\bf Formation of phonon vortex modes via symmetry-adapted linear combinations of degenerate planar modes.} (a) Structure of a hexagonal lattice and the $k$ directions of 3-fold plane waves in (d).
Phonon bands of (b)  a unitcell and (c) a $24\times 24$  supercell.
(d) Six degenerate TA plane waves at $\omega \approx 2.34$ THz.
(e) Six degenerate TA vortex modes at the same frequency, with the winding number of each vortex and the $C_n$ rotational symmetry of the mode indicated. The crosses indicate the sites of $P_{\rm d}^{\rm max}$ (maximum amplitude). 
}
\label{Fig.1}
\end{figure*}

{\bf Introduction}.---
The vibrational properties of two-dimensional (2D) materials with impurities or defects constitute an emerging research topic, enabled by recent advances in atomic-scale visualization of vibrational modes using scanning transmission electron microscopy~\cite{bao2024phonon, jiang2024atomic, cheng2025revealing, jiang2025single, shi2026first}, as well as the visualization of particle motions in 2D dusty plasma crystals~\cite{lo2025multiscale}. 
{\it Phonon vortices} are observed at heavy impurities based on density-functional theory calculations~\cite{bao2024phonon} (see Fig.~\ref{Fig.1}e for examples of phonon vortices).
Such phonon vortices can influence the thermal transport and other properties of the material. However, their origin remains elusive. A heuristic analogy has been proposed, likening a phonon vortex to water flowing around an obstacle~\cite{bao2024phonon}, but this picture is questionable, since a vibrational mode should behave fundamentally differently from a fluid flow.

A different line of research is motivated by the attempt to connect topological defects in the vibrational modes of amorphous solids (glasses) to their structural defects and plastic response to shear~\cite{wu2023topology, wu2025geometry, huang2025spotting, huang2025geometric, vaibhav2025experimental, bera2025hedgehog, baggioli2026topological}. The basic assumption is that ``soft spots'' in the amorphous particle configuration, which are more prone to rearrangement upon mechanical perturbations such as shear, can be identified by topological defects in the vibrational modes. However, this interpretation has been questioned, since phonon vortices---which also look like ``topological defects''---can emerge in pure phonon modes and are obviously not correlated with any structural defects or plasticity~\cite{mizuno2021computational,huang2025geometric}. In general, phonon vortices can hybridize with non-phononic topological defects that truly originate from soft spots \cite{lerner2021low,gartner2016nonlinear}. To distinguish between these, a method based on the Nye vector has been proposed to filter out the latter category~\cite{huang2025geometric}. Furthermore, a particle-level decomposition of phononic and non-phononic components has been proposed using a spatial coherence parameter~\cite{miao2025vibrational}.

Despite these efforts, the current understanding remains largely phenomenological. A key question that needs to be answered is: if ``topological defects'' in vibrational modes indeed correspond to some form of defects in the material, then what is the counterpart behavior of a defect-free perfect crystal, and what happens when a single defect is introduced into such a crystal? Surprisingly, even in this perturbative regime, a first-principles understanding is still lacking. This gap constitutes one of the primary motivations for the present study.

In this work, we combine numerical simulations, group theory, and perturbation theory to develop a first-principles description of phonon vortex formation. Our results lead to several key insights, some of which stand in marked contrast to the conventional understanding.

\textit{(i) Origin of phonon vortices.} We first show that phonon vortices are formed as one possible solution (another possible  solution is planar modes) of the phonon eigenvalue equation in perfect 2D crystals, via the superposition of degenerate planar modes. The superposition coefficients are fully determined by a group-theory derivation based on the lattice symmetry. Thus, phonon vortices do not  generically originate from impurities or defects.

\textit{(ii) Preference and shift of phonon vortices by a heavy impurity.} We then add one heavy impurity (mass defect) to a perfect crystal and show that the vortex modes are favored over planar modes through the combined principle of energy minimization and mode orthogonality. The same principle also governs the spatial shift of the vortex modes and thus the location of the phonon vortices. Consequently, the impurity can coincide with either a nodal point (phonon vortex), or a maximum-amplitude site (not a phonon vortex) that results in localized resonance. Hence, there is no one-to-one correspondence between the location of the impurity and that of a phonon vortex.

\textit{(iii) Suppression of phonon vortices by shear and its competition with the impurity.} Under a shear deformation, phonon vortices in a perfect crystal are suppressed and planar modes become preferred, as shear breaks the rotational symmetry. In the presence of both impurity and shear, their competition determines the crossover between the vortex-dominated and planar-dominated regimes, the boundary of which is predicted by perturbation theory. Thus, shear tends to suppress phonon vortices, rather than giving rise to a plastic event at the heavy impurity site.

{\bf Origin of phonon vortices in perfect crystals: 
symmetry-adapted linear combinations
of degenerate planar modes}---We first show that phonon vortices in perfect crystals arise from linear combinations of degenerate planar modes. The superposition coefficients are uniquely determined by the irreducible representations of the lattice point group.

As a representative example, we consider a standard model of graphene (Gr) consisting of carbon atoms interacting via the AIREBO potential \cite{stuart2000reactive} in a 2D hexagonal lattice with a \(24 \times 24\) supercell (see Fig.~\ref{Fig.1}a).  The phonon eigenvalue equation (see Appendix A) is solved numerically to obtain the phonon bands and normal modes. 

Figures~\ref{Fig.1}(b) and (c) show the phonon bands of the unit cell and supercell, respectively. Owing to Brillouin zone folding, new modes appear at the \(\overline{\Gamma}\) point for the supercell, where the wave vector \(\mathbf{q}=0\). The two lowest frequencies of these new modes are \(\omega \approx 2.34 \) THz for the transverse acoustic (TA) modes and {\(\omega \approx 4.05\)} THz for the longitudinal acoustic (LA) modes (we exclude  Goldstone modes at $\omega = 0$ and out-of-plane  modes in our discussion). Since the hexagonal lattice possesses a 6-fold rotational symmetry (\(C_6\)), the new modes at the \(\overline{\Gamma}\) point exhibit 6-fold degeneracy. Without loss of generality, our discussion below focuses on the six lowest-frequency TA modes at the $\overline{\Gamma}$ point. 

The planar solutions of the six lowest-frequency TA modes are shown in Fig.~\ref{Fig.1}(d). They are divided into two groups (sub-basis); the modes in the two groups differ respectively by a phase shift: { $\varphi_i^{\rm I} (\mathbf{r}) = \hat{\mathbf{e}}_i \exp(i \mathbf{k}_i \cdot \mathbf{r}) $; \(\varphi_i^{\rm II} (\mathbf{r}) = \hat{\mathbf{e}}_i \exp[i(\mathbf{k}_i\cdot \mathbf{r} + \frac{\pi}{2})]\), for \(i=1,2,3 \). 
Here, \(\hat{\mathbf{n}}_1=(1,0), \hat{\mathbf{n}}_2=(-\frac{1}{2}, \frac{\sqrt{3}}{2})\), and \(\hat{\mathbf{n}}_3=(-\frac{1}{2}, -\frac{\sqrt{3}}{2}) \) are unit vectors, 
$\hat{\mathbf{e}}_i=\hat{\mathbf{z}}\times \hat{\mathbf{n}}_i$ are the TA polarization vectors ($\hat{\mathbf{e}}_1=(0,1)$, $\hat{\mathbf{e}}_2=(-\frac{\sqrt{3}}{2},-\frac{1}{2})$, and $\hat{\mathbf{e}}_3=(\frac{\sqrt{3}}{2},-\frac{1}{2})$), and $k=\frac{2\pi}{\lambda}$ is the wave-vector magnitude.
The largest wave-length is $\lambda_{\rm max} = L$ with $L$ the linear size of the supercell. 
}

Because these plane waves are 6-fold degenerate, they can be linearly combined to form a new 6-fold orthonormal basis. Numerical solutions of the phonon eigenvalue equation under periodic boundary conditions, however, show that this linear combination is not arbitrary; rather, the resulting modes clearly display \emph{phonon vortices}, as shown in Fig.~\ref{Fig.1}(e). Since phonon vortices emerge in perfect crystals, their origin cannot be attributed to impurities or disorder. Below, we show that they arise as a consequence of the lattice symmetry.

As shown in Fig.~\ref{Fig.1}(e), the six vortex modes can be also divided into two groups. The vortex modes in each group is  linear combination of the planar modes in the corresponding group: $V^{\alpha}_i = \sum_j c_{ij}^\alpha \varphi^\alpha_j$, where $\alpha = {\rm I}$ or ${\rm II}$.
The coefficients $c_{ij}^\alpha$ are listed in Table.~\ref{tab:1}. Note that the vortex modes $V^{\alpha}_i$ are standing waves.

The coefficients \(c_{ij}^\alpha\) are the symmetry-adapted linear combination (SALC) coefficients, which can be derived from the point group symmetry. 
For each sub-basis (I or II) of planar modes $\varphi^\alpha_i$, the  point group is \(C_3=\{E,R,R^2\}\), where \(E\) denotes the identity element and \(R\) represents a rotation by \(2\pi/3\).
The character table of \(C_3\) (Table~\ref{tab:3}) gives linearly-combined modes that are adapted to the \(C_3\) symmetry (see Appendix B): $\phi_1^\alpha = \frac{1}{\sqrt{3}}(\varphi_1^\alpha+\varphi_2^\alpha+\varphi_3^\alpha)$, $\phi_2^\alpha = \frac{1}{\sqrt{3}}(\varphi_1^\alpha+ \omega \varphi_2^\alpha+ \omega^2\varphi_3^\alpha)$ and $\phi_3^\alpha = \frac{1}{\sqrt{3}}(\varphi_1^\alpha+ \omega^2 \varphi_2^\alpha+ \omega \varphi_3^\alpha)$, where \(\omega=e^{2\pi i/3}\).

\newlength{\colAwidth}
\setlength{\colAwidth}{0.5cm}  

\begin{table}[tb]
  \caption{SALC coefficients  $c_{ij}^\alpha$ of the six lowest frequency vortex modes.
  }
  \label{tab:1}
  \renewcommand{\arraystretch}{1.25}
  \setlength{\tabcolsep}{7pt}
  \begin{ruledtabular}
    \small
    \begin{tabular}{w{c}{\colAwidth} w{c}{\colAwidth} w{c}{\colAwidth} c}
        & $\varphi_1^{\rm I}$ &  $\varphi_2^{\rm I}$ &  $\varphi_3^{\rm I}$ \\
      $V_1^{\rm I}$ & $1/\sqrt{3}$ & $-1/\sqrt{3}$ & $1/\sqrt{3}$ \\
      $V_2^{\rm I}$  & $\sqrt{2/3}$ & $1/\sqrt{6}$ & $-1/\sqrt{6}$ \\
      $V_3^{\rm I}$  & $0$ & $1/\sqrt{2}$ & $1/\sqrt{2}$ \\
      \midrule
        &  $\varphi_1^{\rm II}$ &  $\varphi_2^{\rm II}$ &  $\varphi_3^{\rm II}$ \\
      $V_1^{\rm II}$ & $-1/\sqrt{3}$ & $-1/\sqrt{3}$ & $1/\sqrt{3}$ \\
      $V_2^{\rm II}$  & $\sqrt{2/3}$ & $-1/\sqrt{6}$ & $1/\sqrt{6}$ \\
      $V_3^{\rm II}$ & $0$ & $-1/\sqrt{2}$ & $-1/\sqrt{2}$ \\
    \end{tabular}
  \end{ruledtabular}
\end{table}

Two additional steps are required to obtain the coefficients in Table~\ref{tab:1}.
First, while $\phi_1^\alpha$ is fully symmetric and real, $\phi_2^\alpha$ and $\phi_3^\alpha$ are complex, with \(\phi_2^\alpha=(\phi_3^{\alpha})^*\).
To obtain real orthonormal modes, we  combine them as:
\(\tilde{\phi}_1^\alpha \equiv \phi_1^\alpha \),
\(\tilde{\phi}_2^\alpha \equiv (\phi_2^\alpha+\phi_3^\alpha)/\sqrt{2} \), and
\(\tilde{\phi}_3^\alpha \equiv (\phi_2^\alpha - \phi_3^\alpha) / i\sqrt{2} \).
Second, $\tilde{\phi}_i^{\rm I}$ and $\tilde{\phi}_i^{\rm II}$ are not orthogonal (\(i=1,2,3\)).
To construct an orthonormal basis of six modes, we exploit the sign freedom of the coefficients (for example, both $\varphi_i^\alpha$ and $-\varphi_i^\alpha$ are valid solutions of the phonon eigenvalue equation).
In other words, while the magnitudes of the coefficients in $\tilde{\phi}_i^{\rm I}$ are fixed, their signs may be varied to satisfy the orthonormality condition for all six modes: \(\sum_{n=1}^3 c_{in}c_{jn}=0 \) for $i \neq j$.
This procedure yields the final SALC coefficients of $V_i^\alpha$ in Table~\ref{tab:1}.

Since the formation of phonon vortices originates from lattice symmetry, it is robust against many factors, including the potential, lattice type, and mode frequency and type (see SM Secs.~1-4). The same SALC principle also applies to low-frequency modes in amorphous solids when quasi-localized excitations are absent (see Appendix C). However, we find that the formation of phonon vortices depends crucially on impurities and shear deformations, which we discuss in detail below.

\begin{figure}[tbh]
\centering
\includegraphics[width=\columnwidth]{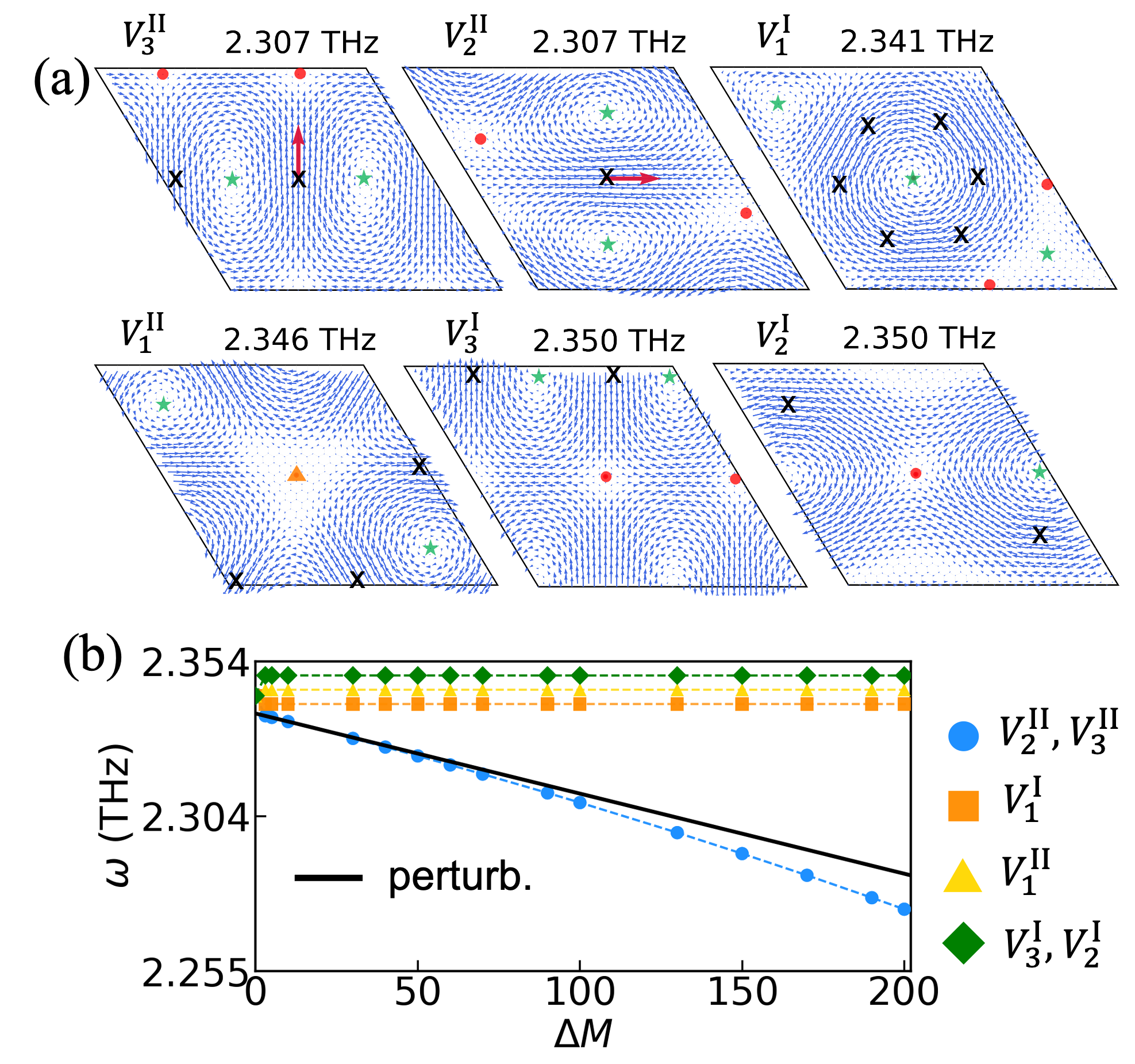}
\caption{Phonon vortex modes with a single heavy mass defect. 
(a) Six lowest-frequency TA modes for $\Delta M = 108$ ($M^{\prime}=120$ u) in a $24$X supercell. The red arrows indicate resonance (see Fig.~\ref{Fig.1}e legend for the meaning of other symbols).
(b) The frequency $\omega$ as a function of $\Delta M$.  The solid line indicates the prediction from the perturbation theory, Eq.~(\ref{eq:impurity}), with $c \approx 2.5 \times 10^{-4}$. }
\label{Fig.2}
\end{figure}

\setlength{\colAwidth}{1cm}

\begin{table}[hb]
  \caption{Parameters obtained by the perturbation theory: $P_{\rm d}^{\rm max}$ and $P_{\rm d}$ in unit of $1/N$, $c$ in unit of ${\rm THz}/u$, and $g$ in unit  of ${\rm THz}$.
  }
  \label{tab:2}
  \renewcommand{\arraystretch}{1.25}
  \setlength{\tabcolsep}{10pt}

  \begin{ruledtabular}
    \small
    \begin{tabular}{
      w{c}{\colAwidth}
      w{c}{1.cm}
      w{c}{1.0cm}
      w{c}{1.0cm}
      w{c}{1.0cm}
    }

      & $P_{\rm d}^{\rm max}$
      & $P_{\rm d}$
      & $c$ 
      & $g$     
      \\

      \midrule
      vortex
      & 
      &  \\
      $V_1^{\rm I}$ & $2.2$  & $0$   & $0$                  & $ - $ \\
      $V_2^{\rm I}$ & $2.5$  & $0$   & $0$                  & $ - $ \\
      $V_3^{\rm I}$ & $3.0$  & $0$   & $0$                  & $ - $ \\
      $V_1^{\rm II}$ & $2.6$ & $0$   & $0$                  & $ - $ \\
      $V_2^{\rm II}$ & $3.0$ & $3.0$ & $2.5 \times 10^{-4}$ & $ - $\\
      $V_3^{\rm II}$ & $3.0$ & $3.0$ & $2.5 \times 10^{-4}$ & $ - $\\

      \midrule

      planar 
      & 
      &  \\
      $\varphi_1^{\rm I}$  & $2.0$ & $ - $ & $1.7 \times 10^{-4}$ & $-7.4$\\   
      $\varphi_2^{\rm I}$  & $2.0$ & $ - $ & $1.7 \times 10^{-4}$ & $-8.2$\\
      $\varphi_3^{\rm I}$  & $2.0$ & $ - $ & $1.7 \times 10^{-4}$ & $ 5.0$\\    
      $\varphi_1^{\rm II}$ & $2.0$ & $ - $ & $1.7 \times 10^{-4}$ & $-7.4$\\
      $\varphi_2^{\rm II}$ & $2.0$ & $ - $ & $1.7 \times 10^{-4}$ & $-8.2$\\
      $\varphi_3^{\rm II}$ & $2.0$ & $ - $ & $1.7 \times 10^{-4}$ & $ 5.0$\\
    \end{tabular}
  \end{ruledtabular}
\end{table}

\begin{figure}[tbh]
\centering
\includegraphics[width=\columnwidth]{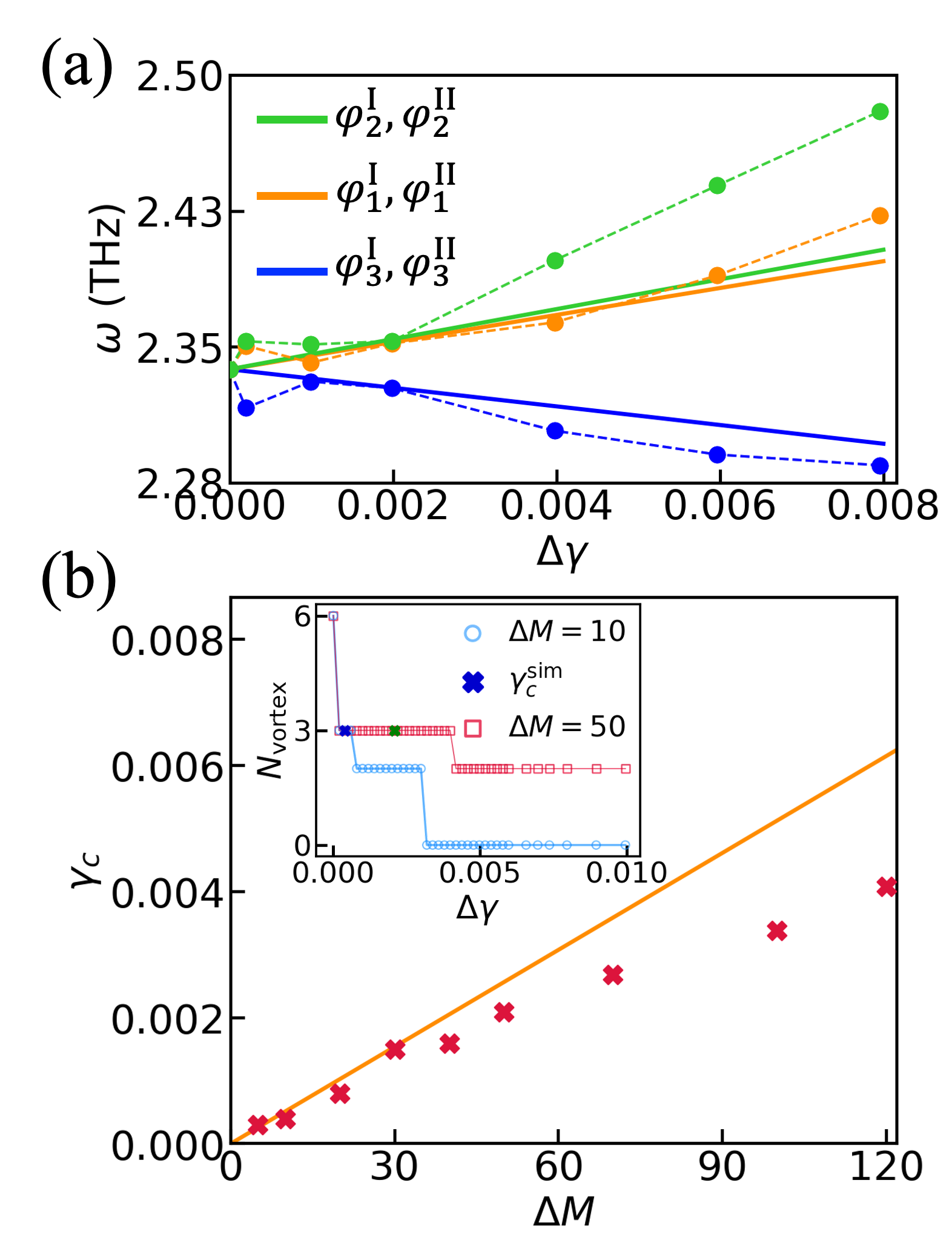}
\caption{Shear-mass competition. (a) Numerical frequency $\omega$ (points) of the six lowest-frequency TA modes, which are all planar, as a function of $\Delta \gamma$ (with a fixed $\Delta M = 108$), compared with the perturbation theory (lines). 
(b) Numerical strain threshold $\gamma_{\rm c}$ as a function of $\Delta M$, compared to Eq.~(\ref{eq:gammac}) (line). (inset) $N_{\rm vortex}$ vs. $\Delta \gamma$ for two given $\Delta M$.}
\label{Fig.3}
\end{figure}
{\bf Preference of phonon vortices for a single heavy impurity}.---
Next, we consider the effects of a single heavy impurity with mass \(M' = M + \Delta M\), where \(M = 12 \, \text{u}\) is the mass of a host carbon atom and \(\Delta M > 0\). In experiments, this corresponds to, for example,  adding a Si impurity in graphene~\cite{bao2024phonon}. 
We model the impurity as a mass defect, neglecting changes to its interactions with other atoms, as these are not central to the present investigation \cite{barker1975optical_1D_chain}.
Without loss of generality, we place the impurity at the center of the supercell; the analysis also applies to a random impurity location. 

Numerically, we find that a single heavy impurity introduces the following effects.
{\it (i) Preference for the vortex modes.} While the six planar modes and the six vortex modes are equivalent in the perfect crystal due to degeneracy, the impurity partially lifts this degeneracy and favors the six vortex modes (see Fig.~\ref{Fig.2}a).
{\it (ii) Shift of modes depending on the impurity location.} The modes are spatially shifted so that the impurity coincides with a high-symmetry point of each mode, 
either a nodal point (zero amplitude) or a point of maximum amplitude.
{\it (iii) Emergence of two resonance modes.} Among the six modes, resonance occurs at the impurity site for two of them (indicated by red arrows in Fig.~\ref{Fig.2}a), namely \(V_2^{\mathrm{II}}\) and \(V_3^{\mathrm{II}}\), for which the impurity coincides with the point of maximum magnitude. Such resonance modes with localized vibration have been observed in the 1D chain model~\cite{barker1975optical_1D_chain} and in recent experiments on 2D materials with heavy impurities~\cite{xu2023single}. Note that for light impurities (\(\Delta M < 0\)), resonant (or locally enhanced) modes appear only at high frequencies~\cite{barker1975optical_1D_chain}. The following analysis does not apply to light impurities, which will be investigated in future studies.

The above numerical observations (i-iii) can be understood from an energetic perspective using a  perturbation theory. The theory yields the frequency shift induced by a heavy impurity (see Appendix D),
\begin{equation}
\omega(\Delta M) \approx \omega(0) - c \Delta M,
\label{eq:impurity}
\end{equation} 
where
\begin{equation}
\label{eq:c}
c = \frac{\omega(0) P_{\rm d}}{2M}, \qquad 
P_{\rm d}= \frac{ \left|\phi^{k}_{(0)} \right|^2}{\sum_{j=1}^N \left|\phi^{j}_{(0)} \right|^2}.
\end{equation}
Here, $\phi_{(0)}$ and $\omega(0)$ denote the eigenvector and the associated eigenfrequency of the unperturbed (defect-free) mode, respectively, while $\phi^{j}_{(0)}$ is the eigenvector component at site $j$. The parameter $P_{\rm d}$ quantifies the intensity weight
of the impurity site $k$ (after the spatial shift) in the unperturbed mode. 
Note that $c \geq 0$, and thus the frequency shifts
downwards.

Next, we apply the above perturbation-theory results to the six lowest-frequency TA modes.
Based on an energetic principle, the impurity system should select a combination that minimizes the energy, i.e., maximizes $c$. Since these modes are degenerate (they share the same $\omega(0)$), maximizing $c$  is equivalent to maximizing $P_{\rm d}$. To this end, we first compute the maximum single-site intensity weight, 
$P_{\rm d}^{\rm max}= {\rm argmax}_{k' \in (1,N)}\left|\phi^{k'}_{(0)} \right|^2/\sum_{j=1}^N \left|\phi^{j}_{(0)} \right|^2$,
for each of the twelve modes ($\varphi_i^\alpha$ and $V_i^\alpha$) shown in Fig.~\ref{Fig.1}(d,e). The site $k_{\rm max}$ that yields the $P_{\rm d}^{\rm max}$ is precisely the one with the largest amplitude in the unperturbed mode (indicated by the cross mark in Fig.~\ref{Fig.1}e). As shown in Table~\ref{tab:2}, among the twelve modes, three of them ($V_3^{\rm I}$, $V_2^{\rm II}$ and $V_3^{\rm II}$) exhibit the largest $P_{\rm d}^{\rm max} \approx 3/N$.


Consequently, any two of the three modes with the largest $P_{\rm d}^{\rm max}$ are selected. 
These two modes are spatially shifted so that the impurity site $k$ coincides $k_{\rm max}$; thus for these two modes, $P_{\rm d}=P_{\rm d}^{\rm max}$.
The negative frequency shift ($\delta \omega = \omega(\delta M)-\omega(0) < 0$) is then achieved via the localized resonance at the impurity site (indicated by the red arrows in Fig.~\ref{Fig.2}a). Because a single impurity site in 2D supports at most two orthogonal vibrational directions, only two of these three highest-$P_{\rm d}^{\rm max}$ modes can be chosen.

Since the two selected modes are vortex modes, the remaining four must also be vortex modes to preserve orthogonality across the full set. Consequently, for each of these four modes, the impurity must be placed at a nodal point where the unperturbed mode amplitude vanishes (which is easy to see if one assumes that the single-site resonance dominates the full mode). As a result, $P_{\rm d} = c  = 0$ for these four modes (see Table~\ref{tab:2}).

The above theory reveals that the degeneracy of the  six vortex modes are partially lifted, with $c \approx 2.5 \times 10^{-4} $ for the two resonant modes, and $c = 0$ for the other four non-resonant modes. This prediction is consistent with the numerical results in Fig.~\ref{Fig.2}(b). The small difference among the four non-resonant modes is due to higher-order corrections. The perturbation analysis thus fully explains our numerical observations (i-iii).

{\bf Suppression  of phonon vortices by  shear deformations}.---The preference for planar modes over vortex modes under a small shear strain $\Delta \gamma$ also follows from symmetry: shear strain breaks rotational invariance, rendering the vortex modes---which exhibit $C_n$ symmetry (Fig.~\ref{Fig.1}e)---incompatible. This is confirmed by perturbation analysis. Expanding the Hessian of the sheared system as $H \approx H_{(0)} + H_\gamma \Delta \gamma$, where $H_\gamma = \left.\frac{d H}{d \gamma}\right|_{\gamma=0}$, we find numerically that $H_\gamma$ is diagonalized by the six planar modes, i.e., $(\varphi_i^\alpha)^T H_\gamma \varphi_j^\beta \neq 0$ only when $i=j$ and $\alpha = \beta$. However, $H_\gamma$ is not diagonalized by the six vortex modes. Thus, the planar modes---and only the planar modes---are eigenfunctions of both $H_\gamma$ and $H$.

The perturbation theory gives the frequency shift of the planar modes:
\begin{equation}
\omega(\Delta \gamma) \approx \omega(0) - g \Delta \gamma,
\label{eq:shear}
\end{equation}
where
\begin{equation}
g = - \left(\frac{1}{2 \omega(0) M} \right) \frac{\phi^T_{(0)} H_\gamma \phi_{(0)}}{\phi^T_{(0)} \phi_{(0)}}.
\end{equation} 
Substituting $\phi_{(0)}$ by the six planar modes yields the $g$ values listed in Table~\ref{tab:2}. The predictions from the perturbation theory are close to  the numerical results for $\omega(\Delta \gamma)$ shown in Fig.~\ref{Fig.3}(a).
The suppression of vortex modes due to breaking rotational symmetry also applies to stretching and compression, which are more experimentally feasible.

{\bf Competition between the impurity and shear.}---
When both the impurity and shear are present, they compete. The impurity favors vortex modes, as these have lower energy according to Eq.~(\ref{eq:impurity}), whereas shear favors planar modes according to Eq.~(\ref{eq:shear}). The crossover from the impurity-dominated vortex regime ($\Delta \gamma < \gamma_{\rm c}$) to the shear-dominated planar regime ($\Delta \gamma > \gamma_{\rm c}$) occurs when the two effects balance ($N_{\mathrm{vortex}} : N_{\mathrm{plane}} = 3 : 3$), yielding the strain threshold:
\begin{equation}
\gamma_{\rm c} = (c/g)\Delta M .
\label{eq:gammac}
\end{equation}
Note that we take $c = 2.5 \times 10^{-4}$ and $g = 5.0$, which correspond to the lowest-energy modes, respectively.

To examine Eq.~(\ref{eq:gammac}), we compute the number of vortex modes $N_{\rm vortex}$ for given values of $\Delta M$ and $\Delta \gamma$. As shown in the inset of Fig.~\ref{Fig.3}(b), for a fixed $\Delta M$, $N_{\rm vortex}=6$ in the absence of shear ($\Delta\gamma=0$) and decreases with increasing $\Delta\gamma$. We define the midpoint of the plateau at $N_{\rm vortex}=3$ as the numerical threshold $\gamma_{\rm c}^{\rm sim}$. The resulting values of $\gamma_{\rm c}^{\rm sim}(\Delta M)$ show good agreement with the perturbative prediction of Eq.~(\ref{eq:gammac}) in the small-$\Delta M$ regime (see Fig.~\ref{Fig.3}b).

\textbf{Discussion.}---When multiple heavy impurities are present,  no simple rules are identified, as the energy-minimizing solution would depend on their specific locations  (see Appendix E). 
The present analysis is restricted to heavy impurities whose behavior is dominated by resonance; the extension of this framework to light impurities, structural defects (dislocations, disclinations, and Stone--Wales defects)~\cite{huang2025spotting}, and soft spots~\cite{lerner2021low} is left for future studies. Another interesting direction concerns phonon vortices in 3D~\cite{wu2025geometry, bera2025hedgehog}. Practically, the defect-shear competition can be used as a tuning approach to engineer the vibrational, thermal, and mechanical properties of the material.\\

\textbf{Acknowledgments}. 
We thank Yang Fu for useful discussions. The authors acknowledge funding from National Key R\&D Program of China
(Grant No. 2025YFF0512000), the China Manned Space
Program (Grant No. CMSS-2025-5-P-002), Wenzhou Institute (No. WIUCASICTP2022) and the National Natural Science Foundation of China (No. 12447101, No. 12404290).
The authors acknowledge the use of the High Performance Cluster at Institute of Theoretical Physics, Chinese Academy of Sciences.

\clearpage

\begin{thebibliography}{36}%
\makeatletter
\providecommand \@ifxundefined [1]{%
 \@ifx{#1\undefined}
}%
\providecommand \@ifnum [1]{%
 \ifnum #1\expandafter \@firstoftwo
 \else \expandafter \@secondoftwo
 \fi
}%
\providecommand \@ifx [1]{%
 \ifx #1\expandafter \@firstoftwo
 \else \expandafter \@secondoftwo
 \fi
}%
\providecommand \natexlab [1]{#1}%
\providecommand \enquote  [1]{``#1''}%
\providecommand \bibnamefont  [1]{#1}%
\providecommand \bibfnamefont [1]{#1}%
\providecommand \citenamefont [1]{#1}%
\providecommand \href@noop [0]{\@secondoftwo}%
\providecommand \href [0]{\begingroup \@sanitize@url \@href}%
\providecommand \@href[1]{\@@startlink{#1}\@@href}%
\providecommand \@@href[1]{\endgroup#1\@@endlink}%
\providecommand \@sanitize@url [0]{\catcode `\\12\catcode `\$12\catcode `\&12\catcode `\#12\catcode `\^12\catcode `\_12\catcode `\%12\relax}%
\providecommand \@@startlink[1]{}%
\providecommand \@@endlink[0]{}%
\providecommand \url  [0]{\begingroup\@sanitize@url \@url }%
\providecommand \@url [1]{\endgroup\@href {#1}{\urlprefix }}%
\providecommand \urlprefix  [0]{URL }%
\providecommand \Eprint [0]{\href }%
\providecommand \doibase [0]{https://doi.org/}%
\providecommand \selectlanguage [0]{\@gobble}%
\providecommand \bibinfo  [0]{\@secondoftwo}%
\providecommand \bibfield  [0]{\@secondoftwo}%
\providecommand \translation [1]{[#1]}%
\providecommand \BibitemOpen [0]{}%
\providecommand \bibitemStop [0]{}%
\providecommand \bibitemNoStop [0]{.\EOS\space}%
\providecommand \EOS [0]{\spacefactor3000\relax}%
\providecommand \BibitemShut  [1]{\csname bibitem#1\endcsname}%
\let\auto@bib@innerbib\@empty
\bibitem [{\citenamefont {Bao}\ \emph {et~al.}(2024)\citenamefont {Bao}, \citenamefont {Xu}, \citenamefont {Li}, \citenamefont {Su}, \citenamefont {Zhou},\ and\ \citenamefont {Pantelides}}]{bao2024phonon}%
  \BibitemOpen
  \bibfield  {author} {\bibinfo {author} {\bibfnamefont {D.-L.}\ \bibnamefont {Bao}}, \bibinfo {author} {\bibfnamefont {M.}~\bibnamefont {Xu}}, \bibinfo {author} {\bibfnamefont {A.-W.}\ \bibnamefont {Li}}, \bibinfo {author} {\bibfnamefont {G.}~\bibnamefont {Su}}, \bibinfo {author} {\bibfnamefont {W.}~\bibnamefont {Zhou}},\ and\ \bibinfo {author} {\bibfnamefont {S.~T.}\ \bibnamefont {Pantelides}},\ }\href@noop {} {\bibfield  {journal} {\bibinfo  {journal} {Nanoscale Horizons}\ }\textbf {\bibinfo {volume} {9}},\ \bibinfo {pages} {248} (\bibinfo {year} {2024})}\BibitemShut {NoStop}%
\bibitem [{\citenamefont {Jiang}\ \emph {et~al.}(2024)\citenamefont {Jiang}, \citenamefont {Wang}, \citenamefont {Zhang}, \citenamefont {Liu}, \citenamefont {Shi}, \citenamefont {Sheng}, \citenamefont {Sheng}, \citenamefont {Ge}, \citenamefont {Wang}, \citenamefont {Shen} \emph {et~al.}}]{jiang2024atomic}%
  \BibitemOpen
  \bibfield  {author} {\bibinfo {author} {\bibfnamefont {H.}~\bibnamefont {Jiang}}, \bibinfo {author} {\bibfnamefont {T.}~\bibnamefont {Wang}}, \bibinfo {author} {\bibfnamefont {Z.}~\bibnamefont {Zhang}}, \bibinfo {author} {\bibfnamefont {F.}~\bibnamefont {Liu}}, \bibinfo {author} {\bibfnamefont {R.}~\bibnamefont {Shi}}, \bibinfo {author} {\bibfnamefont {B.}~\bibnamefont {Sheng}}, \bibinfo {author} {\bibfnamefont {S.}~\bibnamefont {Sheng}}, \bibinfo {author} {\bibfnamefont {W.}~\bibnamefont {Ge}}, \bibinfo {author} {\bibfnamefont {P.}~\bibnamefont {Wang}}, \bibinfo {author} {\bibfnamefont {B.}~\bibnamefont {Shen}}, \emph {et~al.},\ }\href@noop {} {\bibfield  {journal} {\bibinfo  {journal} {Nature communications}\ }\textbf {\bibinfo {volume} {15}},\ \bibinfo {pages} {9052} (\bibinfo {year} {2024})}\BibitemShut {NoStop}%
\bibitem [{\citenamefont {Cheng}\ \emph {et~al.}(2025)\citenamefont {Cheng}, \citenamefont {Jiang}, \citenamefont {Zhang}, \citenamefont {Pi}, \citenamefont {Yang},\ and\ \citenamefont {Deng}}]{cheng2025revealing}%
  \BibitemOpen
  \bibfield  {author} {\bibinfo {author} {\bibfnamefont {M.}~\bibnamefont {Cheng}}, \bibinfo {author} {\bibfnamefont {X.}~\bibnamefont {Jiang}}, \bibinfo {author} {\bibfnamefont {H.}~\bibnamefont {Zhang}}, \bibinfo {author} {\bibfnamefont {X.}~\bibnamefont {Pi}}, \bibinfo {author} {\bibfnamefont {D.}~\bibnamefont {Yang}},\ and\ \bibinfo {author} {\bibfnamefont {T.}~\bibnamefont {Deng}},\ }\href@noop {} {\bibfield  {journal} {\bibinfo  {journal} {Applied Physics Letters}\ }\textbf {\bibinfo {volume} {127}} (\bibinfo {year} {2025})}\BibitemShut {NoStop}%
\bibitem [{\citenamefont {Jiang}\ \emph {et~al.}(2025)\citenamefont {Jiang}, \citenamefont {Wang}, \citenamefont {Zhang}, \citenamefont {Shi}, \citenamefont {Xu}, \citenamefont {Wang}, \citenamefont {Ma}, \citenamefont {Sheng}, \citenamefont {Liu}, \citenamefont {Ge} \emph {et~al.}}]{jiang2025single}%
  \BibitemOpen
  \bibfield  {author} {\bibinfo {author} {\bibfnamefont {H.}~\bibnamefont {Jiang}}, \bibinfo {author} {\bibfnamefont {T.}~\bibnamefont {Wang}}, \bibinfo {author} {\bibfnamefont {Z.}~\bibnamefont {Zhang}}, \bibinfo {author} {\bibfnamefont {R.}~\bibnamefont {Shi}}, \bibinfo {author} {\bibfnamefont {X.}~\bibnamefont {Xu}}, \bibinfo {author} {\bibfnamefont {Z.}~\bibnamefont {Wang}}, \bibinfo {author} {\bibfnamefont {C.}~\bibnamefont {Ma}}, \bibinfo {author} {\bibfnamefont {B.}~\bibnamefont {Sheng}}, \bibinfo {author} {\bibfnamefont {F.}~\bibnamefont {Liu}}, \bibinfo {author} {\bibfnamefont {W.}~\bibnamefont {Ge}}, \emph {et~al.},\ }\href@noop {} {\bibfield  {journal} {\bibinfo  {journal} {Nano Letters}\ }\textbf {\bibinfo {volume} {25}},\ \bibinfo {pages} {14279} (\bibinfo {year} {2025})}\BibitemShut {NoStop}%
\bibitem [{\citenamefont {Shi}\ \emph {et~al.}(2026)\citenamefont {Shi}, \citenamefont {Wang}, \citenamefont {Tan}, \citenamefont {Zheng}, \citenamefont {Chen},\ and\ \citenamefont {Zhang}}]{shi2026first}%
  \BibitemOpen
  \bibfield  {author} {\bibinfo {author} {\bibfnamefont {L.}~\bibnamefont {Shi}}, \bibinfo {author} {\bibfnamefont {X.}~\bibnamefont {Wang}}, \bibinfo {author} {\bibfnamefont {Y.}~\bibnamefont {Tan}}, \bibinfo {author} {\bibfnamefont {Y.}~\bibnamefont {Zheng}}, \bibinfo {author} {\bibfnamefont {Y.}~\bibnamefont {Chen}},\ and\ \bibinfo {author} {\bibfnamefont {Q.}~\bibnamefont {Zhang}},\ }\href@noop {} {\bibfield  {journal} {\bibinfo  {journal} {Solid State Communications}\ ,\ \bibinfo {pages} {116400}} (\bibinfo {year} {2026})}\BibitemShut {NoStop}%
\bibitem [{\citenamefont {Lo}\ \emph {et~al.}(2025)\citenamefont {Lo}, \citenamefont {Siao}, \citenamefont {Zhao},\ and\ \citenamefont {I}}]{lo2025multiscale}%
  \BibitemOpen
  \bibfield  {author} {\bibinfo {author} {\bibfnamefont {W.-S.}\ \bibnamefont {Lo}}, \bibinfo {author} {\bibfnamefont {C.-Y.}\ \bibnamefont {Siao}}, \bibinfo {author} {\bibfnamefont {Y.-C.}\ \bibnamefont {Zhao}},\ and\ \bibinfo {author} {\bibfnamefont {L.}~\bibnamefont {I}},\ }\href@noop {} {\bibfield  {journal} {\bibinfo  {journal} {Physical Review Research}\ }\textbf {\bibinfo {volume} {7}},\ \bibinfo {pages} {L022078} (\bibinfo {year} {2025})}\BibitemShut {NoStop}%
\bibitem [{\citenamefont {Wu}\ \emph {et~al.}(2023)\citenamefont {Wu}, \citenamefont {Chen}, \citenamefont {Wang}, \citenamefont {Kob},\ and\ \citenamefont {Xu}}]{wu2023topology}%
  \BibitemOpen
  \bibfield  {author} {\bibinfo {author} {\bibfnamefont {Z.~W.}\ \bibnamefont {Wu}}, \bibinfo {author} {\bibfnamefont {Y.}~\bibnamefont {Chen}}, \bibinfo {author} {\bibfnamefont {W.-H.}\ \bibnamefont {Wang}}, \bibinfo {author} {\bibfnamefont {W.}~\bibnamefont {Kob}},\ and\ \bibinfo {author} {\bibfnamefont {L.}~\bibnamefont {Xu}},\ }\href@noop {} {\bibfield  {journal} {\bibinfo  {journal} {Nature Communications}\ }\textbf {\bibinfo {volume} {14}},\ \bibinfo {pages} {2955} (\bibinfo {year} {2023})}\BibitemShut {NoStop}%
\bibitem [{\citenamefont {Wu}\ \emph {et~al.}(2025)\citenamefont {Wu}, \citenamefont {Barrat},\ and\ \citenamefont {Kob}}]{wu2025geometry}%
  \BibitemOpen
  \bibfield  {author} {\bibinfo {author} {\bibfnamefont {Z.~W.}\ \bibnamefont {Wu}}, \bibinfo {author} {\bibfnamefont {J.-L.}\ \bibnamefont {Barrat}},\ and\ \bibinfo {author} {\bibfnamefont {W.}~\bibnamefont {Kob}},\ }\href@noop {} {\bibfield  {journal} {\bibinfo  {journal} {Nature Communications}\ } (\bibinfo {year} {2025})}\BibitemShut {NoStop}%
\bibitem [{\citenamefont {Huang}\ \emph {et~al.}(2025{\natexlab{a}})\citenamefont {Huang}, \citenamefont {Wang}, \citenamefont {Jiang},\ and\ \citenamefont {Baggioli}}]{huang2025spotting}%
  \BibitemOpen
  \bibfield  {author} {\bibinfo {author} {\bibfnamefont {L.-Z.}\ \bibnamefont {Huang}}, \bibinfo {author} {\bibfnamefont {Y.-J.}\ \bibnamefont {Wang}}, \bibinfo {author} {\bibfnamefont {M.-Q.}\ \bibnamefont {Jiang}},\ and\ \bibinfo {author} {\bibfnamefont {M.}~\bibnamefont {Baggioli}},\ }\href@noop {} {\bibfield  {journal} {\bibinfo  {journal} {Journal of the Mechanics and Physics of Solids}\ ,\ \bibinfo {pages} {106274}} (\bibinfo {year} {2025}{\natexlab{a}})}\BibitemShut {NoStop}%
\bibitem [{\citenamefont {Huang}\ \emph {et~al.}(2025{\natexlab{b}})\citenamefont {Huang}, \citenamefont {Yang}, \citenamefont {Jiang}, \citenamefont {Wang},\ and\ \citenamefont {Baggioli}}]{huang2025geometric}%
  \BibitemOpen
  \bibfield  {author} {\bibinfo {author} {\bibfnamefont {L.-Z.}\ \bibnamefont {Huang}}, \bibinfo {author} {\bibfnamefont {X.}~\bibnamefont {Yang}}, \bibinfo {author} {\bibfnamefont {M.-Q.}\ \bibnamefont {Jiang}}, \bibinfo {author} {\bibfnamefont {Y.-J.}\ \bibnamefont {Wang}},\ and\ \bibinfo {author} {\bibfnamefont {M.}~\bibnamefont {Baggioli}},\ }\href@noop {} {\bibfield  {journal} {\bibinfo  {journal} {arXiv preprint arXiv:2512.12668}\ } (\bibinfo {year} {2025}{\natexlab{b}})}\BibitemShut {NoStop}%
\bibitem [{\citenamefont {Vaibhav}\ \emph {et~al.}(2025)\citenamefont {Vaibhav}, \citenamefont {Bera}, \citenamefont {Liu}, \citenamefont {Baggioli}, \citenamefont {Keim},\ and\ \citenamefont {Zaccone}}]{vaibhav2025experimental}%
  \BibitemOpen
  \bibfield  {author} {\bibinfo {author} {\bibfnamefont {V.}~\bibnamefont {Vaibhav}}, \bibinfo {author} {\bibfnamefont {A.}~\bibnamefont {Bera}}, \bibinfo {author} {\bibfnamefont {A.~C.}\ \bibnamefont {Liu}}, \bibinfo {author} {\bibfnamefont {M.}~\bibnamefont {Baggioli}}, \bibinfo {author} {\bibfnamefont {P.}~\bibnamefont {Keim}},\ and\ \bibinfo {author} {\bibfnamefont {A.}~\bibnamefont {Zaccone}},\ }\href@noop {} {\bibfield  {journal} {\bibinfo  {journal} {Nature Communications}\ }\textbf {\bibinfo {volume} {16}},\ \bibinfo {pages} {55} (\bibinfo {year} {2025})}\BibitemShut {NoStop}%
\bibitem [{\citenamefont {Bera}\ \emph {et~al.}(2025)\citenamefont {Bera}, \citenamefont {Zaccone},\ and\ \citenamefont {Baggioli}}]{bera2025hedgehog}%
  \BibitemOpen
  \bibfield  {author} {\bibinfo {author} {\bibfnamefont {A.}~\bibnamefont {Bera}}, \bibinfo {author} {\bibfnamefont {A.}~\bibnamefont {Zaccone}},\ and\ \bibinfo {author} {\bibfnamefont {M.}~\bibnamefont {Baggioli}},\ }\href@noop {} {\bibfield  {journal} {\bibinfo  {journal} {Nature Communications}\ }\textbf {\bibinfo {volume} {16}},\ \bibinfo {pages} {5990} (\bibinfo {year} {2025})}\BibitemShut {NoStop}%
\bibitem [{\citenamefont {Baggioli}\ \emph {et~al.}(2026)\citenamefont {Baggioli}, \citenamefont {Falk},\ and\ \citenamefont {Kob}}]{baggioli2026topological}%
  \BibitemOpen
  \bibfield  {author} {\bibinfo {author} {\bibfnamefont {M.}~\bibnamefont {Baggioli}}, \bibinfo {author} {\bibfnamefont {M.~L.}\ \bibnamefont {Falk}},\ and\ \bibinfo {author} {\bibfnamefont {W.}~\bibnamefont {Kob}},\ }\href@noop {} {\bibfield  {journal} {\bibinfo  {journal} {arXiv preprint arXiv:2604.07061}\ } (\bibinfo {year} {2026})}\BibitemShut {NoStop}%
\bibitem [{\citenamefont {Mizuno}\ and\ \citenamefont {Ikeda}(2021)}]{mizuno2021computational}%
  \BibitemOpen
  \bibfield  {author} {\bibinfo {author} {\bibfnamefont {H.}~\bibnamefont {Mizuno}}\ and\ \bibinfo {author} {\bibfnamefont {A.}~\bibnamefont {Ikeda}},\ }\href@noop {} {\bibfield  {journal} {\bibinfo  {journal} {Low-Temperature Thermal and Vibrational Properties of Disordered Solids: A Half-Century of Universal ``Anomalie'' of Glasses}\ ,\ \bibinfo {pages} {375}} (\bibinfo {year} {2021})}\BibitemShut {NoStop}%
\bibitem [{\citenamefont {Lerner}\ and\ \citenamefont {Bouchbinder}(2021)}]{lerner2021low}%
  \BibitemOpen
  \bibfield  {author} {\bibinfo {author} {\bibfnamefont {E.}~\bibnamefont {Lerner}}\ and\ \bibinfo {author} {\bibfnamefont {E.}~\bibnamefont {Bouchbinder}},\ }\href@noop {} {\bibfield  {journal} {\bibinfo  {journal} {The Journal of chemical physics}\ }\textbf {\bibinfo {volume} {155}} (\bibinfo {year} {2021})}\BibitemShut {NoStop}%
\bibitem [{\citenamefont {Gartner}\ and\ \citenamefont {Lerner}(2016)}]{gartner2016nonlinear}%
  \BibitemOpen
  \bibfield  {author} {\bibinfo {author} {\bibfnamefont {L.}~\bibnamefont {Gartner}}\ and\ \bibinfo {author} {\bibfnamefont {E.}~\bibnamefont {Lerner}},\ }\href@noop {} {\bibfield  {journal} {\bibinfo  {journal} {SciPost Physics}\ }\textbf {\bibinfo {volume} {1}},\ \bibinfo {pages} {016} (\bibinfo {year} {2016})}\BibitemShut {NoStop}%
\bibitem [{\citenamefont {Miao}\ \emph {et~al.}(2025)\citenamefont {Miao}, \citenamefont {Ivlev}, \citenamefont {L{\"o}wen}, \citenamefont {Nosenko}, \citenamefont {Huang}, \citenamefont {Yang}, \citenamefont {Thomas}, \citenamefont {Zhang},\ and\ \citenamefont {Du}}]{miao2025vibrational}%
  \BibitemOpen
  \bibfield  {author} {\bibinfo {author} {\bibfnamefont {Y.}~\bibnamefont {Miao}}, \bibinfo {author} {\bibfnamefont {A.~V.}\ \bibnamefont {Ivlev}}, \bibinfo {author} {\bibfnamefont {H.}~\bibnamefont {L{\"o}wen}}, \bibinfo {author} {\bibfnamefont {V.}~\bibnamefont {Nosenko}}, \bibinfo {author} {\bibfnamefont {H.}~\bibnamefont {Huang}}, \bibinfo {author} {\bibfnamefont {W.}~\bibnamefont {Yang}}, \bibinfo {author} {\bibfnamefont {H.~M.}\ \bibnamefont {Thomas}}, \bibinfo {author} {\bibfnamefont {J.}~\bibnamefont {Zhang}},\ and\ \bibinfo {author} {\bibfnamefont {C.-R.}\ \bibnamefont {Du}},\ }\href@noop {} {\bibfield  {journal} {\bibinfo  {journal} {Physical Review Letters}\ }\textbf {\bibinfo {volume} {135}},\ \bibinfo {pages} {135301} (\bibinfo {year} {2025})}\BibitemShut {NoStop}%
\bibitem [{\citenamefont {Stuart}\ \emph {et~al.}(2000)\citenamefont {Stuart}, \citenamefont {Tutein},\ and\ \citenamefont {Harrison}}]{stuart2000reactive}%
  \BibitemOpen
  \bibfield  {author} {\bibinfo {author} {\bibfnamefont {S.~J.}\ \bibnamefont {Stuart}}, \bibinfo {author} {\bibfnamefont {A.~B.}\ \bibnamefont {Tutein}},\ and\ \bibinfo {author} {\bibfnamefont {J.~A.}\ \bibnamefont {Harrison}},\ }\href@noop {} {\bibfield  {journal} {\bibinfo  {journal} {The Journal of chemical physics}\ }\textbf {\bibinfo {volume} {112}},\ \bibinfo {pages} {6472} (\bibinfo {year} {2000})}\BibitemShut {NoStop}%
\bibitem [{\citenamefont {Barker~Jr}\ and\ \citenamefont {Sievers}(1975)}]{barker1975optical_1D_chain}%
  \BibitemOpen
  \bibfield  {author} {\bibinfo {author} {\bibfnamefont {A.}~\bibnamefont {Barker~Jr}}\ and\ \bibinfo {author} {\bibfnamefont {A.~J.}\ \bibnamefont {Sievers}},\ }\href@noop {} {\bibfield  {journal} {\bibinfo  {journal} {Reviews of Modern Physics}\ }\textbf {\bibinfo {volume} {47}},\ \bibinfo {pages} {S1} (\bibinfo {year} {1975})}\BibitemShut {NoStop}%
\bibitem [{\citenamefont {Xu}\ \emph {et~al.}(2023)\citenamefont {Xu}, \citenamefont {Bao}, \citenamefont {Li}, \citenamefont {Gao}, \citenamefont {Meng}, \citenamefont {Li}, \citenamefont {Du}, \citenamefont {Su}, \citenamefont {Pennycook}, \citenamefont {Pantelides} \emph {et~al.}}]{xu2023single}%
  \BibitemOpen
  \bibfield  {author} {\bibinfo {author} {\bibfnamefont {M.}~\bibnamefont {Xu}}, \bibinfo {author} {\bibfnamefont {D.-L.}\ \bibnamefont {Bao}}, \bibinfo {author} {\bibfnamefont {A.}~\bibnamefont {Li}}, \bibinfo {author} {\bibfnamefont {M.}~\bibnamefont {Gao}}, \bibinfo {author} {\bibfnamefont {D.}~\bibnamefont {Meng}}, \bibinfo {author} {\bibfnamefont {A.}~\bibnamefont {Li}}, \bibinfo {author} {\bibfnamefont {S.}~\bibnamefont {Du}}, \bibinfo {author} {\bibfnamefont {G.}~\bibnamefont {Su}}, \bibinfo {author} {\bibfnamefont {S.~J.}\ \bibnamefont {Pennycook}}, \bibinfo {author} {\bibfnamefont {S.~T.}\ \bibnamefont {Pantelides}}, \emph {et~al.},\ }\href@noop {} {\bibfield  {journal} {\bibinfo  {journal} {Nature Materials}\ }\textbf {\bibinfo {volume} {22}},\ \bibinfo {pages} {612} (\bibinfo {year} {2023})}\BibitemShut {NoStop}%
\bibitem [{\citenamefont {Togo}(2023)}]{togo2023first}%
  \BibitemOpen
  \bibfield  {author} {\bibinfo {author} {\bibfnamefont {A.}~\bibnamefont {Togo}},\ }\href@noop {} {\bibfield  {journal} {\bibinfo  {journal} {Journal of the Physical Society of Japan}\ }\textbf {\bibinfo {volume} {92}},\ \bibinfo {pages} {012001} (\bibinfo {year} {2023})}\BibitemShut {NoStop}%
\bibitem [{\citenamefont {Togo}\ and\ \citenamefont {Tanaka}(2015)}]{togo2015first}%
  \BibitemOpen
  \bibfield  {author} {\bibinfo {author} {\bibfnamefont {A.}~\bibnamefont {Togo}}\ and\ \bibinfo {author} {\bibfnamefont {I.}~\bibnamefont {Tanaka}},\ }\href@noop {} {\bibfield  {journal} {\bibinfo  {journal} {Scripta materialia}\ }\textbf {\bibinfo {volume} {108}},\ \bibinfo {pages} {1} (\bibinfo {year} {2015})}\BibitemShut {NoStop}%
\bibitem [{\citenamefont {Dresselhaus}\ \emph {et~al.}(2007)\citenamefont {Dresselhaus}, \citenamefont {Dresselhaus},\ and\ \citenamefont {Jorio}}]{dresselhaus2007group}%
  \BibitemOpen
  \bibfield  {author} {\bibinfo {author} {\bibfnamefont {M.~S.}\ \bibnamefont {Dresselhaus}}, \bibinfo {author} {\bibfnamefont {G.}~\bibnamefont {Dresselhaus}},\ and\ \bibinfo {author} {\bibfnamefont {A.}~\bibnamefont {Jorio}},\ }\href@noop {} {\emph {\bibinfo {title} {Group theory: application to the physics of condensed matter}}}\ (\bibinfo  {publisher} {Springer Science \& Business Media},\ \bibinfo {year} {2007})\BibitemShut {NoStop}%
\bibitem [{\citenamefont {Toh}\ \emph {et~al.}(2020)\citenamefont {Toh}, \citenamefont {Zhang}, \citenamefont {Lin}, \citenamefont {Mayorov}, \citenamefont {Wang}, \citenamefont {Orofeo}, \citenamefont {Ferry}, \citenamefont {Andersen}, \citenamefont {Kakenov}, \citenamefont {Guo} \emph {et~al.}}]{toh2020synthesis_MAC}%
  \BibitemOpen
  \bibfield  {author} {\bibinfo {author} {\bibfnamefont {C.-T.}\ \bibnamefont {Toh}}, \bibinfo {author} {\bibfnamefont {H.}~\bibnamefont {Zhang}}, \bibinfo {author} {\bibfnamefont {J.}~\bibnamefont {Lin}}, \bibinfo {author} {\bibfnamefont {A.~S.}\ \bibnamefont {Mayorov}}, \bibinfo {author} {\bibfnamefont {Y.-P.}\ \bibnamefont {Wang}}, \bibinfo {author} {\bibfnamefont {C.~M.}\ \bibnamefont {Orofeo}}, \bibinfo {author} {\bibfnamefont {D.~B.}\ \bibnamefont {Ferry}}, \bibinfo {author} {\bibfnamefont {H.}~\bibnamefont {Andersen}}, \bibinfo {author} {\bibfnamefont {N.}~\bibnamefont {Kakenov}}, \bibinfo {author} {\bibfnamefont {Z.}~\bibnamefont {Guo}}, \emph {et~al.},\ }\href@noop {} {\bibfield  {journal} {\bibinfo  {journal} {Nature}\ }\textbf {\bibinfo {volume} {577}},\ \bibinfo {pages} {199} (\bibinfo {year} {2020})}\BibitemShut {NoStop}%
\bibitem [{\citenamefont {Bai}\ \emph {et~al.}(2024)\citenamefont {Bai}, \citenamefont {Hu}, \citenamefont {Li}, \citenamefont {Zhang}, \citenamefont {Li}, \citenamefont {Zhang}, \citenamefont {Xue}, \citenamefont {Jiang}, \citenamefont {Wang}, \citenamefont {Cui} \emph {et~al.}}]{bai2024nitrogen_MAC}%
  \BibitemOpen
  \bibfield  {author} {\bibinfo {author} {\bibfnamefont {X.}~\bibnamefont {Bai}}, \bibinfo {author} {\bibfnamefont {P.}~\bibnamefont {Hu}}, \bibinfo {author} {\bibfnamefont {A.}~\bibnamefont {Li}}, \bibinfo {author} {\bibfnamefont {Y.}~\bibnamefont {Zhang}}, \bibinfo {author} {\bibfnamefont {A.}~\bibnamefont {Li}}, \bibinfo {author} {\bibfnamefont {G.}~\bibnamefont {Zhang}}, \bibinfo {author} {\bibfnamefont {Y.}~\bibnamefont {Xue}}, \bibinfo {author} {\bibfnamefont {T.}~\bibnamefont {Jiang}}, \bibinfo {author} {\bibfnamefont {Z.}~\bibnamefont {Wang}}, \bibinfo {author} {\bibfnamefont {H.}~\bibnamefont {Cui}}, \emph {et~al.},\ }\href@noop {} {\bibfield  {journal} {\bibinfo  {journal} {Nature}\ }\textbf {\bibinfo {volume} {634}},\ \bibinfo {pages} {80} (\bibinfo {year} {2024})}\BibitemShut {NoStop}%
\bibitem [{\citenamefont {Lin}\ \emph {et~al.}(2025)\citenamefont {Lin}, \citenamefont {Jiang}, \citenamefont {Dou}, \citenamefont {Lyu}, \citenamefont {Han}, \citenamefont {Meng}, \citenamefont {He}, \citenamefont {Zhou}, \citenamefont {Li}, \citenamefont {Lin} \emph {et~al.}}]{lin2025ultraclean_MAC}%
  \BibitemOpen
  \bibfield  {author} {\bibinfo {author} {\bibfnamefont {H.}~\bibnamefont {Lin}}, \bibinfo {author} {\bibfnamefont {J.}~\bibnamefont {Jiang}}, \bibinfo {author} {\bibfnamefont {Y.}~\bibnamefont {Dou}}, \bibinfo {author} {\bibfnamefont {P.}~\bibnamefont {Lyu}}, \bibinfo {author} {\bibfnamefont {X.}~\bibnamefont {Han}}, \bibinfo {author} {\bibfnamefont {Y.}~\bibnamefont {Meng}}, \bibinfo {author} {\bibfnamefont {Y.}~\bibnamefont {He}}, \bibinfo {author} {\bibfnamefont {X.}~\bibnamefont {Zhou}}, \bibinfo {author} {\bibfnamefont {K.}~\bibnamefont {Li}}, \bibinfo {author} {\bibfnamefont {G.}~\bibnamefont {Lin}}, \emph {et~al.},\ }\href@noop {} {\bibfield  {journal} {\bibinfo  {journal} {Nature Nanotechnology}\ }\textbf {\bibinfo {volume} {20}},\ \bibinfo {pages} {1431} (\bibinfo {year} {2025})}\BibitemShut {NoStop}%
\bibitem [{\citenamefont {Tian}\ \emph {et~al.}(2023)\citenamefont {Tian}, \citenamefont {Ma}, \citenamefont {Li}, \citenamefont {Cheng}, \citenamefont {Ning}, \citenamefont {Han}, \citenamefont {Xu}, \citenamefont {Zhang}, \citenamefont {Zhao}, \citenamefont {Li} \emph {et~al.}}]{tian2023disorder_MAC}%
  \BibitemOpen
  \bibfield  {author} {\bibinfo {author} {\bibfnamefont {H.}~\bibnamefont {Tian}}, \bibinfo {author} {\bibfnamefont {Y.}~\bibnamefont {Ma}}, \bibinfo {author} {\bibfnamefont {Z.}~\bibnamefont {Li}}, \bibinfo {author} {\bibfnamefont {M.}~\bibnamefont {Cheng}}, \bibinfo {author} {\bibfnamefont {S.}~\bibnamefont {Ning}}, \bibinfo {author} {\bibfnamefont {E.}~\bibnamefont {Han}}, \bibinfo {author} {\bibfnamefont {M.}~\bibnamefont {Xu}}, \bibinfo {author} {\bibfnamefont {P.-F.}\ \bibnamefont {Zhang}}, \bibinfo {author} {\bibfnamefont {K.}~\bibnamefont {Zhao}}, \bibinfo {author} {\bibfnamefont {R.}~\bibnamefont {Li}}, \emph {et~al.},\ }\href@noop {} {\bibfield  {journal} {\bibinfo  {journal} {Nature}\ }\textbf {\bibinfo {volume} {615}},\ \bibinfo {pages} {56} (\bibinfo {year} {2023})}\BibitemShut {NoStop}%
\bibitem [{\citenamefont {Gastellu}\ \emph {et~al.}(2022)\citenamefont {Gastellu}, \citenamefont {Kilgour},\ and\ \citenamefont {Simine}}]{gastellu2022electronic}%
  \BibitemOpen
  \bibfield  {author} {\bibinfo {author} {\bibfnamefont {N.}~\bibnamefont {Gastellu}}, \bibinfo {author} {\bibfnamefont {M.}~\bibnamefont {Kilgour}},\ and\ \bibinfo {author} {\bibfnamefont {L.}~\bibnamefont {Simine}},\ }\href@noop {} {\bibfield  {journal} {\bibinfo  {journal} {The Journal of Physical Chemistry Letters}\ }\textbf {\bibinfo {volume} {13}},\ \bibinfo {pages} {339} (\bibinfo {year} {2022})}\BibitemShut {NoStop}%
\bibitem [{\citenamefont {Felix}\ \emph {et~al.}(2020)\citenamefont {Felix}, \citenamefont {Tromer}, \citenamefont {Autreto}, \citenamefont {Ribeiro~Junior},\ and\ \citenamefont {Galvao}}]{felix2020mechanical}%
  \BibitemOpen
  \bibfield  {author} {\bibinfo {author} {\bibfnamefont {L.~C.}\ \bibnamefont {Felix}}, \bibinfo {author} {\bibfnamefont {R.~M.}\ \bibnamefont {Tromer}}, \bibinfo {author} {\bibfnamefont {P.~A.}\ \bibnamefont {Autreto}}, \bibinfo {author} {\bibfnamefont {L.~A.}\ \bibnamefont {Ribeiro~Junior}},\ and\ \bibinfo {author} {\bibfnamefont {D.~S.}\ \bibnamefont {Galvao}},\ }\href@noop {} {\bibfield  {journal} {\bibinfo  {journal} {The Journal of Physical Chemistry C}\ }\textbf {\bibinfo {volume} {124}},\ \bibinfo {pages} {14855} (\bibinfo {year} {2020})}\BibitemShut {NoStop}%
\bibitem [{\citenamefont {Garz{\'o}n-Ram{\'\i}rez}\ \emph {et~al.}(2022)\citenamefont {Garz{\'o}n-Ram{\'\i}rez}, \citenamefont {Gastellu},\ and\ \citenamefont {Simine}}]{garzon2022optoelectronic}%
  \BibitemOpen
  \bibfield  {author} {\bibinfo {author} {\bibfnamefont {A.~J.}\ \bibnamefont {Garz{\'o}n-Ram{\'\i}rez}}, \bibinfo {author} {\bibfnamefont {N.}~\bibnamefont {Gastellu}},\ and\ \bibinfo {author} {\bibfnamefont {L.}~\bibnamefont {Simine}},\ }\href@noop {} {\bibfield  {journal} {\bibinfo  {journal} {The Journal of Physical Chemistry Letters}\ }\textbf {\bibinfo {volume} {13}},\ \bibinfo {pages} {1057} (\bibinfo {year} {2022})}\BibitemShut {NoStop}%
\bibitem [{\citenamefont {Xie}\ and\ \citenamefont {Wei}(2021)}]{xie2021roughening}%
  \BibitemOpen
  \bibfield  {author} {\bibinfo {author} {\bibfnamefont {W.}~\bibnamefont {Xie}}\ and\ \bibinfo {author} {\bibfnamefont {Y.}~\bibnamefont {Wei}},\ }\href@noop {} {\bibfield  {journal} {\bibinfo  {journal} {Nano Letters}\ }\textbf {\bibinfo {volume} {21}},\ \bibinfo {pages} {4823} (\bibinfo {year} {2021})}\BibitemShut {NoStop}%
\bibitem [{\citenamefont {Zhang}\ \emph {et~al.}(2022{\natexlab{a}})\citenamefont {Zhang}, \citenamefont {Wang}, \citenamefont {Zhang}, \citenamefont {Zhang}, \citenamefont {Du},\ and\ \citenamefont {Pantelides}}]{zhang2022structure_maBN}%
  \BibitemOpen
  \bibfield  {author} {\bibinfo {author} {\bibfnamefont {Y.-T.}\ \bibnamefont {Zhang}}, \bibinfo {author} {\bibfnamefont {Y.-P.}\ \bibnamefont {Wang}}, \bibinfo {author} {\bibfnamefont {X.}~\bibnamefont {Zhang}}, \bibinfo {author} {\bibfnamefont {Y.-Y.}\ \bibnamefont {Zhang}}, \bibinfo {author} {\bibfnamefont {S.}~\bibnamefont {Du}},\ and\ \bibinfo {author} {\bibfnamefont {S.~T.}\ \bibnamefont {Pantelides}},\ }\href@noop {} {\bibfield  {journal} {\bibinfo  {journal} {Nano Letters}\ }\textbf {\bibinfo {volume} {22}},\ \bibinfo {pages} {8018} (\bibinfo {year} {2022}{\natexlab{a}})}\BibitemShut {NoStop}%
\bibitem [{\citenamefont {Zhang}\ \emph {et~al.}(2022{\natexlab{b}})\citenamefont {Zhang}, \citenamefont {Wang}, \citenamefont {Zhang}, \citenamefont {Du},\ and\ \citenamefont {Pantelides}}]{zhang2022thermal_maBN}%
  \BibitemOpen
  \bibfield  {author} {\bibinfo {author} {\bibfnamefont {Y.-T.}\ \bibnamefont {Zhang}}, \bibinfo {author} {\bibfnamefont {Y.-P.}\ \bibnamefont {Wang}}, \bibinfo {author} {\bibfnamefont {Y.-Y.}\ \bibnamefont {Zhang}}, \bibinfo {author} {\bibfnamefont {S.}~\bibnamefont {Du}},\ and\ \bibinfo {author} {\bibfnamefont {S.~T.}\ \bibnamefont {Pantelides}},\ }\href@noop {} {\bibfield  {journal} {\bibinfo  {journal} {Applied Physics Letters}\ }\textbf {\bibinfo {volume} {120}} (\bibinfo {year} {2022}{\natexlab{b}})}\BibitemShut {NoStop}%
\bibitem [{\citenamefont {Zhang}\ \emph {et~al.}(2024)\citenamefont {Zhang}, \citenamefont {Zhang}, \citenamefont {Wang}, \citenamefont {Li}, \citenamefont {Du}, \citenamefont {Zhang},\ and\ \citenamefont {Pantelides}}]{zhang2024structural_mechanical}%
  \BibitemOpen
  \bibfield  {author} {\bibinfo {author} {\bibfnamefont {X.}~\bibnamefont {Zhang}}, \bibinfo {author} {\bibfnamefont {Y.-T.}\ \bibnamefont {Zhang}}, \bibinfo {author} {\bibfnamefont {Y.-P.}\ \bibnamefont {Wang}}, \bibinfo {author} {\bibfnamefont {S.}~\bibnamefont {Li}}, \bibinfo {author} {\bibfnamefont {S.}~\bibnamefont {Du}}, \bibinfo {author} {\bibfnamefont {Y.-Y.}\ \bibnamefont {Zhang}},\ and\ \bibinfo {author} {\bibfnamefont {S.~T.}\ \bibnamefont {Pantelides}},\ }\href@noop {} {\bibfield  {journal} {\bibinfo  {journal} {Physical Review B}\ }\textbf {\bibinfo {volume} {109}},\ \bibinfo {pages} {174106} (\bibinfo {year} {2024})}\BibitemShut {NoStop}%
\bibitem [{\citenamefont {Zhuang}\ \emph {et~al.}(2016)\citenamefont {Zhuang}, \citenamefont {Zhao}, \citenamefont {Dong}, \citenamefont {Yan},\ and\ \citenamefont {Ding}}]{zhuang2016evolution_kMC}%
  \BibitemOpen
  \bibfield  {author} {\bibinfo {author} {\bibfnamefont {J.}~\bibnamefont {Zhuang}}, \bibinfo {author} {\bibfnamefont {R.}~\bibnamefont {Zhao}}, \bibinfo {author} {\bibfnamefont {J.}~\bibnamefont {Dong}}, \bibinfo {author} {\bibfnamefont {T.}~\bibnamefont {Yan}},\ and\ \bibinfo {author} {\bibfnamefont {F.}~\bibnamefont {Ding}},\ }\href@noop {} {\bibfield  {journal} {\bibinfo  {journal} {Physical Chemistry Chemical Physics}\ }\textbf {\bibinfo {volume} {18}},\ \bibinfo {pages} {2932} (\bibinfo {year} {2016})}\BibitemShut {NoStop}%
\bibitem [{\citenamefont {Ding}\ and\ \citenamefont {Yakobson}(2014)}]{ding2014energy_kMC}%
  \BibitemOpen
  \bibfield  {author} {\bibinfo {author} {\bibfnamefont {F.}~\bibnamefont {Ding}}\ and\ \bibinfo {author} {\bibfnamefont {B.~I.}\ \bibnamefont {Yakobson}},\ }\href@noop {} {\bibfield  {journal} {\bibinfo  {journal} {The Journal of Physical Chemistry Letters}\ }\textbf {\bibinfo {volume} {5}},\ \bibinfo {pages} {2922} (\bibinfo {year} {2014})}\BibitemShut {NoStop}%
\end{thebibliography}%


\begin{thebibliography}{7}%
\makeatletter
\providecommand \@ifxundefined [1]{%
 \@ifx{#1\undefined}
}%
\providecommand \@ifnum [1]{%
 \ifnum #1\expandafter \@firstoftwo
 \else \expandafter \@secondoftwo
 \fi
}%
\providecommand \@ifx [1]{%
 \ifx #1\expandafter \@firstoftwo
 \else \expandafter \@secondoftwo
 \fi
}%
\providecommand \natexlab [1]{#1}%
\providecommand \enquote  [1]{``#1''}%
\providecommand \bibnamefont  [1]{#1}%
\providecommand \bibfnamefont [1]{#1}%
\providecommand \citenamefont [1]{#1}%
\providecommand \href@noop [0]{\@secondoftwo}%
\providecommand \href [0]{\begingroup \@sanitize@url \@href}%
\providecommand \@href[1]{\@@startlink{#1}\@@href}%
\providecommand \@@href[1]{\endgroup#1\@@endlink}%
\providecommand \@sanitize@url [0]{\catcode `\\12\catcode `\$12\catcode `\&12\catcode `\#12\catcode `\^12\catcode `\_12\catcode `\%12\relax}%
\providecommand \@@startlink[1]{}%
\providecommand \@@endlink[0]{}%
\providecommand \url  [0]{\begingroup\@sanitize@url \@url }%
\providecommand \@url [1]{\endgroup\@href {#1}{\urlprefix }}%
\providecommand \urlprefix  [0]{URL }%
\providecommand \Eprint [0]{\href }%
\providecommand \doibase [0]{https://doi.org/}%
\providecommand \selectlanguage [0]{\@gobble}%
\providecommand \bibinfo  [0]{\@secondoftwo}%
\providecommand \bibfield  [0]{\@secondoftwo}%
\providecommand \translation [1]{[#1]}%
\providecommand \BibitemOpen [0]{}%
\providecommand \bibitemStop [0]{}%
\providecommand \bibitemNoStop [0]{.\EOS\space}%
\providecommand \EOS [0]{\spacefactor3000\relax}%
\providecommand \BibitemShut  [1]{\csname bibitem#1\endcsname}%
\let\auto@bib@innerbib\@empty
\bibitem [{\citenamefont {Zhang}\ \emph {et~al.}(2026)\citenamefont {Zhang}, \citenamefont {Huang}, \citenamefont {Du}, \citenamefont {Ying}, \citenamefont {Du},\ and\ \citenamefont {Zhang}}]{zhang2026comprehensive}%
  \BibitemOpen
  \bibfield  {author} {\bibinfo {author} {\bibfnamefont {S.}~\bibnamefont {Zhang}}, \bibinfo {author} {\bibfnamefont {Z.}~\bibnamefont {Huang}}, \bibinfo {author} {\bibfnamefont {M.}~\bibnamefont {Du}}, \bibinfo {author} {\bibfnamefont {T.}~\bibnamefont {Ying}}, \bibinfo {author} {\bibfnamefont {L.}~\bibnamefont {Du}},\ and\ \bibinfo {author} {\bibfnamefont {T.}~\bibnamefont {Zhang}},\ }\href@noop {} {\bibfield  {journal} {\bibinfo  {journal} {Physical Review B}\ }\textbf {\bibinfo {volume} {113}},\ \bibinfo {pages} {024302} (\bibinfo {year} {2026})}\BibitemShut {NoStop}%
\bibitem [{\citenamefont {Togo}\ \emph {et~al.}(2023)\citenamefont {Togo}, \citenamefont {Chaput}, \citenamefont {Tadano},\ and\ \citenamefont {Tanaka}}]{togo2023implementation}%
  \BibitemOpen
  \bibfield  {author} {\bibinfo {author} {\bibfnamefont {A.}~\bibnamefont {Togo}}, \bibinfo {author} {\bibfnamefont {L.}~\bibnamefont {Chaput}}, \bibinfo {author} {\bibfnamefont {T.}~\bibnamefont {Tadano}},\ and\ \bibinfo {author} {\bibfnamefont {I.}~\bibnamefont {Tanaka}},\ }\href@noop {} {\bibfield  {journal} {\bibinfo  {journal} {Journal of Physics: Condensed Matter}\ }\textbf {\bibinfo {volume} {35}},\ \bibinfo {pages} {353001} (\bibinfo {year} {2023})}\BibitemShut {NoStop}%
\bibitem [{\citenamefont {Togo}(2023)}]{togo2023first}%
  \BibitemOpen
  \bibfield  {author} {\bibinfo {author} {\bibfnamefont {A.}~\bibnamefont {Togo}},\ }\href@noop {} {\bibfield  {journal} {\bibinfo  {journal} {Journal of the Physical Society of Japan}\ }\textbf {\bibinfo {volume} {92}},\ \bibinfo {pages} {012001} (\bibinfo {year} {2023})}\BibitemShut {NoStop}%
\bibitem [{\citenamefont {Thompson}\ \emph {et~al.}(2022)\citenamefont {Thompson}, \citenamefont {Aktulga}, \citenamefont {Berger}, \citenamefont {Bolintineanu}, \citenamefont {Brown}, \citenamefont {Crozier}, \citenamefont {{In't Veld}}, \citenamefont {Kohlmeyer}, \citenamefont {Moore}, \citenamefont {Nguyen} \emph {et~al.}}]{thompson2022lammps}%
  \BibitemOpen
  \bibfield  {author} {\bibinfo {author} {\bibfnamefont {A.~P.}\ \bibnamefont {Thompson}}, \bibinfo {author} {\bibfnamefont {H.~M.}\ \bibnamefont {Aktulga}}, \bibinfo {author} {\bibfnamefont {R.}~\bibnamefont {Berger}}, \bibinfo {author} {\bibfnamefont {D.~S.}\ \bibnamefont {Bolintineanu}}, \bibinfo {author} {\bibfnamefont {W.~M.}\ \bibnamefont {Brown}}, \bibinfo {author} {\bibfnamefont {P.~S.}\ \bibnamefont {Crozier}}, \bibinfo {author} {\bibfnamefont {P.~J.}\ \bibnamefont {{In't Veld}}}, \bibinfo {author} {\bibfnamefont {A.}~\bibnamefont {Kohlmeyer}}, \bibinfo {author} {\bibfnamefont {S.~G.}\ \bibnamefont {Moore}}, \bibinfo {author} {\bibfnamefont {T.~D.}\ \bibnamefont {Nguyen}}, \emph {et~al.},\ }\href@noop {} {\bibfield  {journal} {\bibinfo  {journal} {Computer physics communications}\ }\textbf {\bibinfo {volume} {271}},\ \bibinfo {pages} {108171} (\bibinfo {year} {2022})}\BibitemShut {NoStop}%
\bibitem [{\citenamefont {Plimpton}(1995)}]{plimpton1995fast}%
  \BibitemOpen
  \bibfield  {author} {\bibinfo {author} {\bibfnamefont {S.}~\bibnamefont {Plimpton}},\ }\href@noop {} {\bibfield  {journal} {\bibinfo  {journal} {Journal of computational physics}\ }\textbf {\bibinfo {volume} {117}},\ \bibinfo {pages} {1} (\bibinfo {year} {1995})}\BibitemShut {NoStop}%
\bibitem [{\citenamefont {Kresse}\ and\ \citenamefont {Furthm{\"u}ller}(1996)}]{kresse1996efficiency}%
  \BibitemOpen
  \bibfield  {author} {\bibinfo {author} {\bibfnamefont {G.}~\bibnamefont {Kresse}}\ and\ \bibinfo {author} {\bibfnamefont {J.}~\bibnamefont {Furthm{\"u}ller}},\ }\href@noop {} {\bibfield  {journal} {\bibinfo  {journal} {Computational materials science}\ }\textbf {\bibinfo {volume} {6}},\ \bibinfo {pages} {15} (\bibinfo {year} {1996})}\BibitemShut {NoStop}%
\bibitem [{\citenamefont {Perdew}\ \emph {et~al.}(1996)\citenamefont {Perdew}, \citenamefont {Burke},\ and\ \citenamefont {Ernzerhof}}]{perdew1996generalized}%
  \BibitemOpen
  \bibfield  {author} {\bibinfo {author} {\bibfnamefont {J.~P.}\ \bibnamefont {Perdew}}, \bibinfo {author} {\bibfnamefont {K.}~\bibnamefont {Burke}},\ and\ \bibinfo {author} {\bibfnamefont {M.}~\bibnamefont {Ernzerhof}},\ }\href@noop {} {\bibfield  {journal} {\bibinfo  {journal} {Physical review letters}\ }\textbf {\bibinfo {volume} {77}},\ \bibinfo {pages} {3865} (\bibinfo {year} {1996})}\BibitemShut {NoStop}%
\end{thebibliography}%


\begin{thebibliography}{0}%
\makeatletter
\providecommand \@ifxundefined [1]{%
 \@ifx{#1\undefined}
}%
\providecommand \@ifnum [1]{%
 \ifnum #1\expandafter \@firstoftwo
 \else \expandafter \@secondoftwo
 \fi
}%
\providecommand \@ifx [1]{%
 \ifx #1\expandafter \@firstoftwo
 \else \expandafter \@secondoftwo
 \fi
}%
\providecommand \natexlab [1]{#1}%
\providecommand \enquote  [1]{``#1''}%
\providecommand \bibnamefont  [1]{#1}%
\providecommand \bibfnamefont [1]{#1}%
\providecommand \citenamefont [1]{#1}%
\providecommand \href@noop [0]{\@secondoftwo}%
\providecommand \href [0]{\begingroup \@sanitize@url \@href}%
\providecommand \@href[1]{\@@startlink{#1}\@@href}%
\providecommand \@@href[1]{\endgroup#1\@@endlink}%
\providecommand \@sanitize@url [0]{\catcode `\\12\catcode `\$12\catcode `\&12\catcode `\#12\catcode `\^12\catcode `\_12\catcode `\%12\relax}%
\providecommand \@@startlink[1]{}%
\providecommand \@@endlink[0]{}%
\providecommand \url  [0]{\begingroup\@sanitize@url \@url }%
\providecommand \@url [1]{\endgroup\@href {#1}{\urlprefix }}%
\providecommand \urlprefix  [0]{URL }%
\providecommand \Eprint [0]{\href }%
\providecommand \doibase [0]{https://doi.org/}%
\providecommand \selectlanguage [0]{\@gobble}%
\providecommand \bibinfo  [0]{\@secondoftwo}%
\providecommand \bibfield  [0]{\@secondoftwo}%
\providecommand \translation [1]{[#1]}%
\providecommand \BibitemOpen [0]{}%
\providecommand \bibitemStop [0]{}%
\providecommand \bibitemNoStop [0]{.\EOS\space}%
\providecommand \EOS [0]{\spacefactor3000\relax}%
\providecommand \BibitemShut  [1]{\csname bibitem#1\endcsname}%
\let\auto@bib@innerbib\@empty
\end{thebibliography}%
\putbib[references]

\clearpage

{
\setlength{\textfloatsep}{5pt}
\setlength{\floatsep}{5pt}
\setlength{\intextsep}{5pt}

\begin{widetext}
\begin{center}
{\Large\bfseries End Matter}
\end{center}
\end{widetext}


\textit{Appendix A. Phonon eigenvalue equation}---
We consider a system within a {\it supercell} (SC), which is a large cell
formed by repeating the original primitive cell (of a linear size $a$) an integer number of times along its lattice vector directions~\cite{togo2023first}. The supercell has periodic boundary conditions (PBC). The supercell defines the degree of freedom of the problem (d.o.f). For a perfect crystal, if the supercell consists of  $L^d$ unit cells in $d$ dimensions, and each unit cell has $n_{\rm a}$ atoms, then the supercell has in total $N = L^d \times n_{\rm a}$ atoms. {It is sufficient to consider only the $x-y$ d.o.f, because the $z$-d.o.f is decoupled from in-plane modes.} Supercells are useful to study effects of impurities and disorder, when the  perfect crystal periodicity breaks down.


The phonon eigenvalue equation is, 
\begin{equation}
D(\mathbf{q}) \phi(\mathbf{q}) = \omega^2(q) \phi(\mathbf{q}),
\label{eq:eigenequation}
\end{equation}
where $\mathbf{q}$  is the wavevector, $\omega$ is the  phonon frequency, and $\phi(\mathbf{q})$  is the corresponding phonon eigenvector.
The dynamical matrix $D$ is  defined as~\cite{togo2015first},
\begin{equation}
D_{kk^{\prime}}^{\alpha \beta}(\mathbf{q})= \sum_{l^{\prime}=0}^\infty\dfrac{H_{\alpha \beta}(0 k,l^{\prime} k^{\prime})}{\sqrt{m_k m_{k^{\prime}}}} e^{(i\mathbf{q}[\mathbf{r}(l^{\prime} k^{\prime})-\mathbf{r}(0 k)])},
\label{eq:D}
\end{equation}
where $\alpha,\beta =  x, y$ are Cartesian indices, $l'$ is the index of repeated supercells, $k=1, 2, \ldots N$ is the atom index, $\mathbf{r}(l^{\prime} k^{\prime})$ is the position of the $k'$-th atom in the $l'$-th supercell, and 
$m_k=M =12$ u is the atom mass  for graphene (Gr).
The Hessian (force constant) matrix is
\begin{equation}
H_{\alpha \beta}(0k,l^{\prime}k')=\dfrac{\partial^2 U}{\partial{u_{0k}^{\alpha}}\partial{u_{l'k'}^{\beta}}}\Bigg|_{u=0},
\end{equation}
where $U$ is the potential energy, and $u$ is the displacement field. The dynamical matrix $D$ is a $dN \times dN$ matrix, and  correspondingly, there are $dN$ eigenvectors ($n=1,2,\ldots, dN$).

In this study, we only study the zero-$q$ (long-wave length) limit, corresponding to the $\Gamma$-point in the SC phonon band. Equations~(\ref{eq:eigenequation}) and~(\ref{eq:D}) then become $q$-independent:
\begin{equation}
D \phi = \omega^2 \phi,
\label{eq:eigenequation2}
\end{equation}
and 
\begin{equation}
D_{kk^{\prime}}^{\alpha \beta}= \sum_{l^{\prime}=0}^\infty\dfrac{H_{\alpha \beta}(0 k,l^{\prime} k^{\prime})}{\sqrt{m_k m_{k^{\prime}}}}.
\label{eq:D2}
\end{equation}

Under harmonic and nearest-neighbor (1NN) approximations, 
\(D=-\dfrac{Kh^2}{m}\nabla^2 \), where $K$ is the spring constant, and $h$ is the lattice spacing. Thus Eq.~(\ref{eq:eigenequation2}) becomes a Helmholtz equation:
\begin{equation}
\nabla^2 \phi = -\dfrac{\omega^2 m}{K h^2} \phi.
\label{eq:harmonic}
\end{equation}
In this continuum limit, crystals and amorphous solids become indistinguishable, as microscopic structures are irrelevant. The solutions are completely determined by the harmonic Eq.~(\ref{eq:harmonic}) and boundary conditions.

\textit{Appendix B. Character table of $C_3$}---
The character table of \(C_3\) is shown in Table~\ref{tab:3}. 
The complex 1D irreducible representations (irreps) have characters \(\chi_m(R^j)=\omega^{mj}\) with \(m=0,1,2\) and \(\omega=e^{2\pi i/3}\). 
The projection operator onto the irreps \(\Gamma_m\) reads \cite{dresselhaus2007group}:
\(
\mathcal P^{(m)} = \frac{d_m}{|G|}\sum_{j=0}^{2} \chi_m(R^j)^*\, D(R^j)
= \frac{1}{3}\sum_{j=0}^2 \omega^{-m j}\,D(R^j).
\)
Here, the dimension of the irreps is \( d_m=1 \), \(|G|=|C_3|=3\) is the order of group, 
and \(D(R)\) is the representation matrix of the rotation \(R\), which acts on the planar modes as follows:
\(D(R)\varphi_1^\alpha = \varphi_2^\alpha\), \(D(R)\varphi_2^\alpha = \varphi_3^\alpha\), and \(D(R)\varphi_3^\alpha = \varphi_1^\alpha\). 
Applying the projection operator \(\mathcal P^{(m)}\) to \(\varphi_1^\alpha\) yields \(\phi_i^\alpha = \mathcal P^{(m)} \varphi_1^\alpha\), which gives the SALC coefficients for the corresponding irrep \(\Gamma_i\), with \(i = m+1\).

\setlength{\colAwidth}{0.5cm}  
\begin{table}[!t]
  \caption{Character table of $C_3$, where \(d_m\) is the dimension of  irreps.}

  \label{tab:3}
  \renewcommand{\arraystretch}{1.25}
  \begin{ruledtabular}
    \setlength{\tabcolsep}{7pt}
    \newlength{\lastcolwidth}
    \setlength{\lastcolwidth}{
      \dimexpr\columnwidth - 4\colAwidth - 10\tabcolsep\relax
    }
    \begin{tabular}{
      w{c}{\colAwidth}
      w{c}{\colAwidth}
      w{c}{\colAwidth}
      w{c}{\colAwidth}
      c
    }
      
      irrep(\(d_m\)) & $E$ & $R$ & $R^2$ & SALC \\
      \midrule
      $\Gamma_1\,(1)$ & $1$ & $1$ & $1$
      & \parbox[t]{\lastcolwidth}{\raggedright
          $\phi_1^\alpha=\dfrac{1}{\sqrt{3}}
          (\varphi_1^\alpha+\varphi_2^\alpha+\varphi_3^\alpha)$
        } \\
      \addlinespace
      $\Gamma_2\,(1)$ & $1$ & $\omega$ & $\omega^2$
      & \parbox[t]{\lastcolwidth}{\raggedright
          $\phi_2^\alpha=\dfrac{1}{\sqrt{3}}
          (\varphi_1^\alpha+\omega\varphi_2^\alpha+\omega^2\varphi_3^\alpha)$
        } \\
      \addlinespace

      $\Gamma_3\,(1)$ & $1$ & $\omega^2$ & $\omega$
      & \parbox[t]{\lastcolwidth}{\raggedright
          $\phi_3^\alpha=\dfrac{1}{\sqrt{3}}
          (\varphi_1^\alpha+\omega^2\varphi_2^\alpha+\omega\varphi_3^\alpha)$
        } \\
      \addlinespace
       \midrule
      {} & {} & {} & {}
      & \begin{tabular}[c]{@{}l@{}}
          $\tilde{\phi}_2^\alpha=\dfrac{1}{\sqrt{6}}
          (2\varphi_1^\alpha-\varphi_2^\alpha-\varphi_3^\alpha)$ \\[4pt]
          $\tilde{\phi}_3^\alpha=\dfrac{1}{\sqrt{2}}
          (\varphi_2^\alpha-\varphi_3^\alpha)$
        \end{tabular} \\
    \end{tabular}
  \end{ruledtabular}
\end{table}

\textit{Appendix C. Monolayer amorphous carbon.}---
We choose monolayer amorphous carbon (MAC) as a model system to study phonon vortices in amorphous solids. MAC is a typical covalently bonded 2D glass that has been experimentally synthesized and studied in recent years \cite{toh2020synthesis_MAC,bai2024nitrogen_MAC,lin2025ultraclean_MAC,tian2023disorder_MAC,gastellu2022electronic,felix2020mechanical,garzon2022optoelectronic,xie2021roughening,zhang2022structure_maBN,zhang2022thermal_maBN,zhang2024structural_mechanical}. Here, we employ a kinetic Monte Carlo algorithm \cite{zhuang2016evolution_kMC,ding2014energy_kMC,zhang2022structure_maBN,zhang2022thermal_maBN,zhang2024structural_mechanical} to numerically prepare MAC configurations.

To examine whether the low-frequency modes remain phononic, we distinguish phononic and non-phononic atoms in a mode $\phi$ based on spatial coherence \cite{miao2025vibrational}:
\[
\delta_i = \frac{1}{n_i} \sum_{j=1}^{n_i} \frac{\phi^i \cdot \phi^j}{|\phi^i| \, |\phi^j|},
\]
where $n_i$ is the number of nearest neighbors of the $i$-th particle. This order parameter measures the angular coherence of neighboring displacement vectors. We then classify an atom as \textit{phononic} if $\delta_i > 0.90$, \textit{non-phononic} if $\delta_i < 0.80$. and \textit{hybrid} otherwise.

Figure~\ref{Fig:MAC} shows that the six lowest-frequency modes are overall extended vortex modes, similar to those in the perfect crystal. Disorder introduces only a small fraction of non-phononic atoms, but the vibrations of these atoms merely distort the phonon vortex modes slightly. No Eshelby-like quasi-localized excitations~\cite{lerner2021low} are observed for the six lowest-frequency modes. Thus, in the long-wavelength limit, amorphous solids and crystals exhibit similar vortex modes, as the microscopic structure becomes irrelevant.

\enlargethispage{2\baselineskip}

\begin{figure}[tbh]
\centering
\includegraphics[width=\columnwidth]{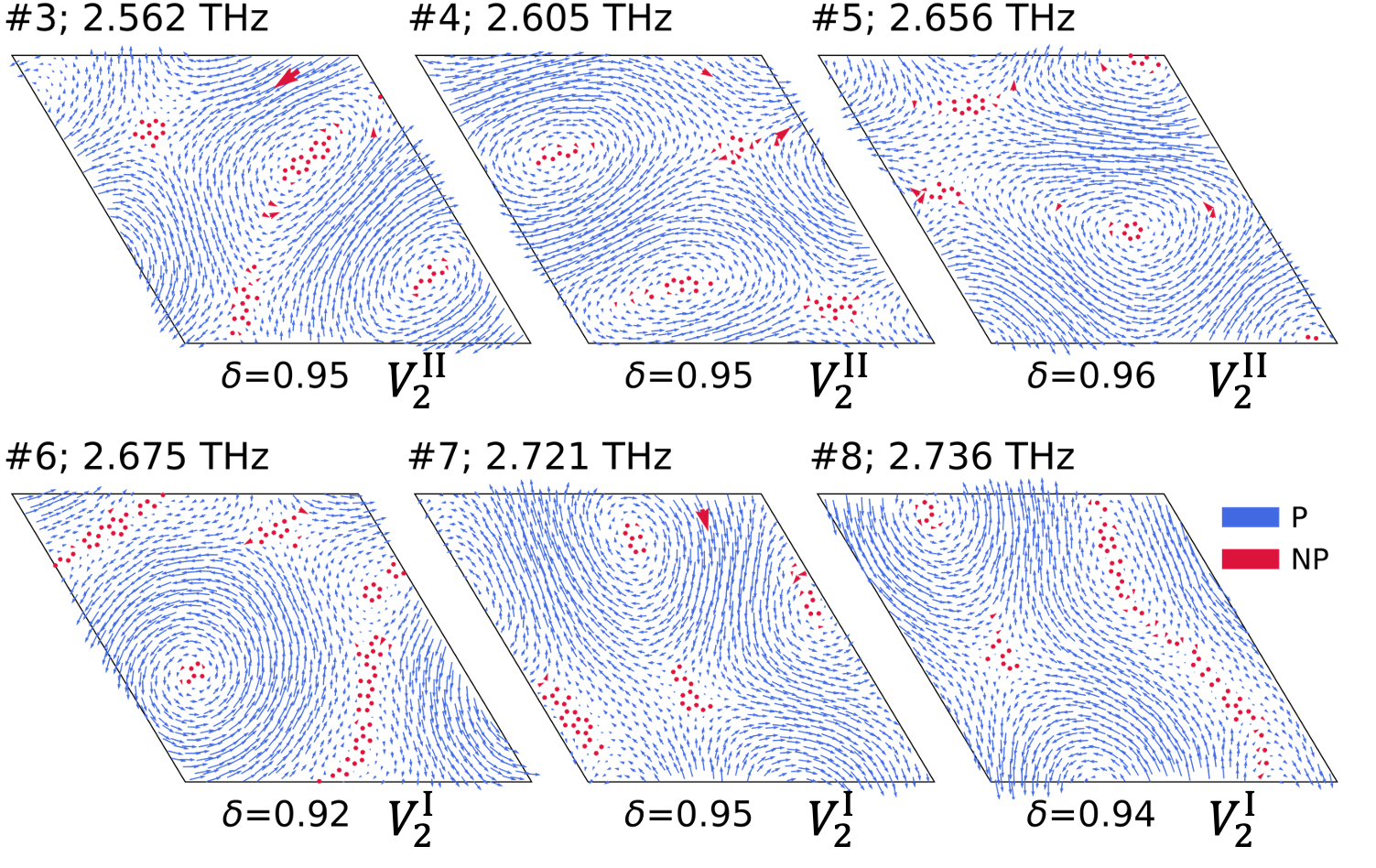}
\caption{Six lowest-frequency modes for a MAC configuration.
The vibrations of phononic (P) and non-phononic (NP) atoms are indicated by blue and red arrows (enlarged for better visualization).
}
\label{Fig:MAC}
\end{figure}

\textit{Appendix D. Perturbation theory for  a single heavy defect.}---
For a heavy defect ($\Delta M > 0$) at a site index $k$, the squared frequency is given by the Rayleigh quotient
\begin{equation}
	\omega^2 = \dfrac{\phi^T H \phi}{\phi^T (M \mathcal{I} + \Delta M \mathcal{P}_k) \phi},
\end{equation}
where $\mathcal{I}$ is the identity matrix and $\mathcal{P}_k$ is the projection operator onto the impurity site $k$, defined by $\mathcal{P}_k \phi^j = \phi^j$ if $j=k$, and $\mathcal{P}_k \phi^j = 0$ otherwise.

To  first order, we replace $\phi$ by the unperturbed mode $\phi_{(0)}$, yielding
\begin{equation}
\begin{split}
	\omega^2 & \approx \dfrac{\phi_{(0)}^T H \phi_{(0)}}{\phi^T_{(0)} (M \mathcal{I} + \Delta M \mathcal{P}_k) \phi_{(0)}} \\
& = \dfrac{\omega^2(0) M}{M \sum_i |\phi^i_{(0)}|^2 + \Delta M |\phi^k_{(0)}|^2},
\end{split}
\label{eq:omega2}
\end{equation}
where we have used $H \phi_{(0)} = \omega^2(0) M \phi_{(0)}$. Simplifying Eq.~(\ref{eq:omega2}) gives Eq.~(\ref{eq:impurity}).

\textit{Appendix E. Multiple heavy impurities.}---When $N_{\rm d}$ heavy mass defects are introduced, the perturbation forms in Eqs.~(\ref{eq:impurity}) and~(\ref{eq:c}) remain valid. However, $P_{\rm d}$ must be modified accordingly as
\begin{equation} 
\label{eq:Pd2}
P_{\rm d}= \frac{ \sum_{i=1}^{N_{\rm d}}\left|\phi^{k_i}_{(0)} \right|^2}{\sum_{j=1}^N \left|\phi^{j}_{(0)} \right|^2},
\end{equation}
where $k_1, k_2, \ldots, k_{N_{\rm d}}$ denote the site indices of the $N_{\rm d}$ impurities, which we assume to be randomly distributed.

For multiple impurities, the modes are shifted, rotated, or recombined so that $P_{\rm d}$ in Eq.~(\ref{eq:Pd2}) is maximized. This optimization, however, depends sensitively on the specific impurity positions. Consequently, no simple general rules exist, in contrast to the single-impurity case. Figure~\ref{Fig:multi_impurity}(a) illustrates an example with 20 randomly placed impurities, where the six lowest-frequency modes emerge as mixtures of planar and vortex modes. 
Interestingly, resonance generally appears at the impurities. 
For a different realization of impurity positions, the composition of these six modes will change accordingly.

For sufficiently large $N_{\rm d}$, we may assume $P_{\rm d} \approx N_{\rm d}/N$. In this case, the perturbation-theory  frequency becomes
\begin{equation}
\label{eq:omega_multi}
\omega = \omega(0) - \dfrac{\omega(0) \Delta M N_{\rm d}}{2 M N}.
\end{equation}
This result is considerably close to the numerically obtained $\omega(N_{\rm d})$, as shown in Fig.~\ref{Fig:multi_impurity}(b), although the frequency splitting among the six modes is not captured by this simple approximation.

No Eshelby-like quasi-localized excitations are observed in the six lowest-frequency modes as $N_{\rm d}$ increases. To demonstrate this, we plot in Fig.~\ref{Fig:multi_impurity}(c) the spatial coherence $\delta = \frac{1}{N} \sum_{i=1}^N \delta_i$,
and the participation ratio 
$PR = \frac{\left(\sum_{i=1}^N \lvert \phi^i \rvert^2\right)^2}{N \sum_{i=1}^N \lvert \phi^i \rvert^4}$, averaged over 
lowest 12 modes (6TA + 6LA).
As $N_{\rm d}$ increases, $\overline{PR}$ decreases due to the growing fraction of resonant atoms; however, $\bar{\delta}$ remains high. This indicates that all six modes are phononic in nature, since quasi-localized excitations would instead reduce $\bar{\delta}$.

\begin{figure}[tbh]
\centering
\includegraphics[width=\columnwidth]{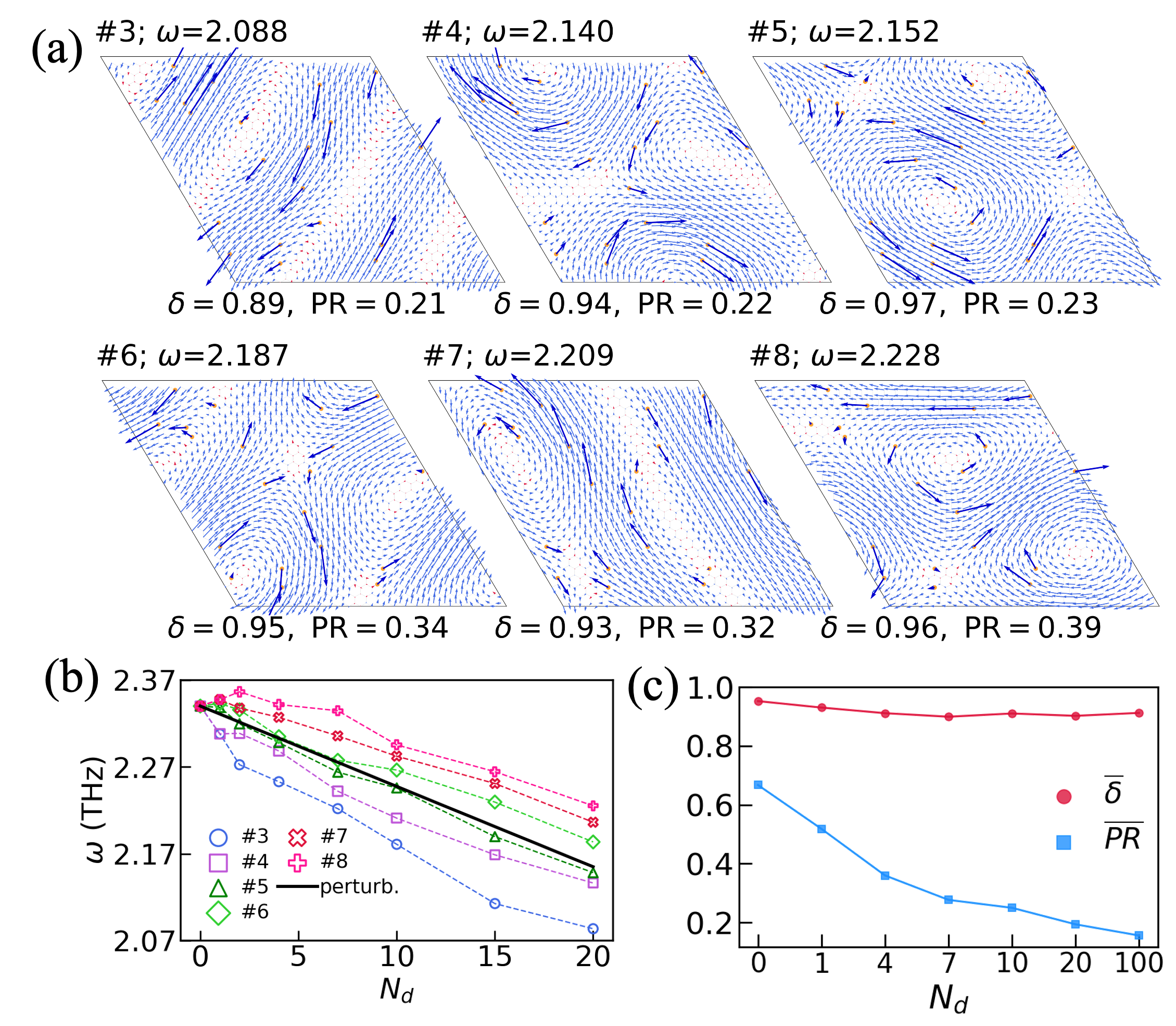}
\caption{Phonon vortex modes with multiple heavy mass defects. 
(a) Six lowest-frequency TA modes for 20 heavy impurities ($\Delta M = 108$)  in a $24 \times 24$ supercell. The dark blue arrows indicate resonance. 
(b) The frequency $\omega$ as a function of $N_{\rm d}$ for the six modes, averaged over multiple random realizations of the impurity positions. The black solid line indicates Eq.~(\ref{eq:omega_multi}). 
(c) The mean participation ratio and spatial coherence as a function of $N_{\rm d}$.
}
\label{Fig:multi_impurity}
\end{figure}

\clearpage

\onecolumngrid
}
\end{bibunit}

\setcounter{figure}{0}
\setcounter{equation}{0}
\setcounter{table}{0}
\setcounter{section}{0}
\renewcommand\thefigure{S\arabic{figure}}
\renewcommand\theequation{S\arabic{equation}}
\renewcommand\thesection{S\arabic{section}}
\renewcommand\thetable{S\arabic{table}}

\begin{bibunit}[apsrev4-2]
\begin{center}
{\Large\bfseries Supplementary  Materials for \\
Emergence and suppression of phonon vortices in two-dimensional crystals: Interplay
of lattice symmetry, heavy impurities, and shear}

\vspace{1em}

Yu-Tian Zhang$^{1,2}$, Deng Pan$^{1,3}$, Yuliang Jin$^{1,2,4,*}$

\vspace{0.5em}

$^1$Institute of Theoretical Physics, Chinese Academy of Sciences, Beijing 100190, China\\
$^2$School of Physical Sciences, University of Chinese Academy of Sciences, Beijing 100049, China\\
$^3$School of Physics and Information Technology, Shaanxi Normal University, Xi’an 710119, China \\
$^4$Center for Theoretical Interdisciplinary Sciences, Wenzhou Institute, University of Chinese Academy of Sciences, Wenzhou, Zhejiang 325001, China\\
$^{*}$Corresponding author: jinyuliang@itp.ac.cn
\end{center}

\vspace{1cm}

\begin{center}
{\large\bfseries CONTENTS}
\end{center}

\vspace{1em}

\smcontentsline{I.}{Phonon vortices in a square lattice}{sec:SM1}

\smcontentsline{II.}{Independence of phonon vortices from interatomic potentials}{sec:SM2}

\smcontentsline{III.}{High frequency phonon vortices}{sec:SM3}

\smcontentsline{IV.}{Fixed boundary conditions}{sec:SM4}

\smcontentsline{}{References}{sec:SMreferences}


\vspace{1cm}

\clearpage

\section{Phonon vortices in a square lattice}
\label{sec:SM1}

\begin{figure}[b!]
\centering
\includegraphics[width=0.65\columnwidth]{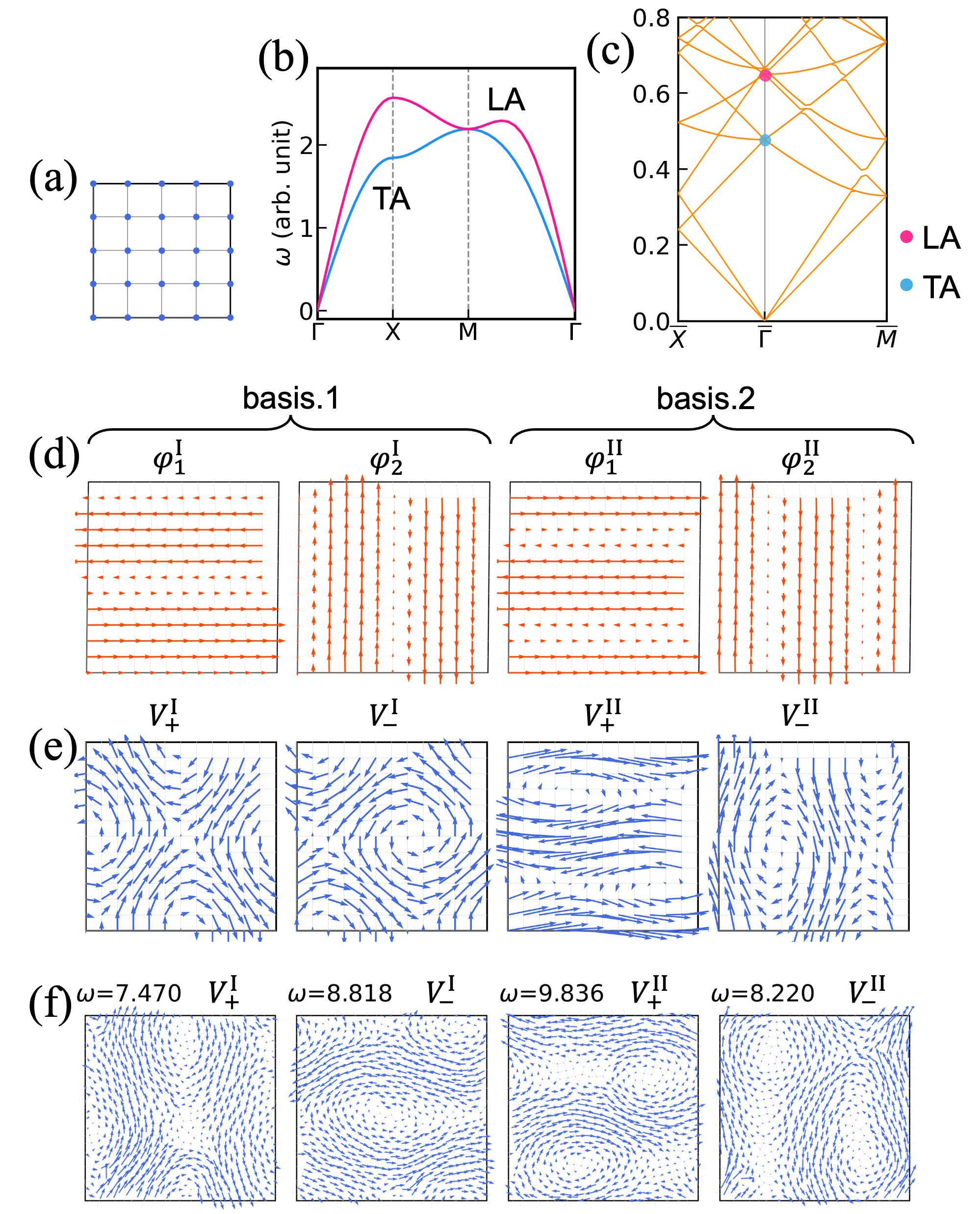}
\caption{Results for a square lattice. (a) Lattice structure. (b) Unitcell phonon band. (c) Phonon band in a $12\times12$ supercell. (d) Four lowest-frequency TA plane waves. (e) Four lowest-frequency TA vortex modes. 
(f) Four lowest-frequency modes for a 2D random network in a supercell of  size $L=34 \mathring{\mathrm{A}}$ (AIREBO interaction).
}
\label{Fig.S1}
\end{figure}

Square lattices support fourfold-degenerate phonon vortices. To model these, we employ a tight-binding (TB) Hamiltonian on a square lattice, incorporating a harmonic potential that includes both longitudinal and transverse interactions \cite{zhang2026comprehensive}.  Here we use $K_1=1.0$, $K_2=0.55$, $T_1=0.2$ and $T_2=0.1$, where $K_n$ are the longitudinal spring constants for the $n$-th neighbors, and $T_n$ are respectively the transverse spring constants.  The lowest four TA planar, or vortex, modes are presented in Fig.~\ref{Fig.S1}.

The fourfold-degenerate plane waves can be separated into two distinct groups, each consisting of two orthogonal plane waves. These groups are distinguished by a relative phase shift of $\pi/2$. They form an effective $C_2$ point group representation.  Within each group, the SALCs are given by the $C_2$ character table, with symmetric combination:
$V_{+}^{\alpha} = (\varphi_1^{\alpha} + \varphi_2^{\alpha})/\sqrt{2}$,
and the antisymmetric combination:
$V_{-}^{\alpha} = (\varphi_1^{\alpha} - \varphi_2^{\alpha})/\sqrt{2}$.

For an amorphous 2D network in a square supercell (Fig.~S2f), the eigenvectors are phonon vortices that resemble those of the square lattice. This validates our SALC principle as a general rule for low-frequency vibrational modes in  glasses, when quasi-localized excitations are absent.

\clearpage

\section{Independence of Phonon Vortices from Interatomic Potentials}
\label{sec:SM2}

\begin{figure*}[htb]
\centering
\includegraphics[width=\textwidth]{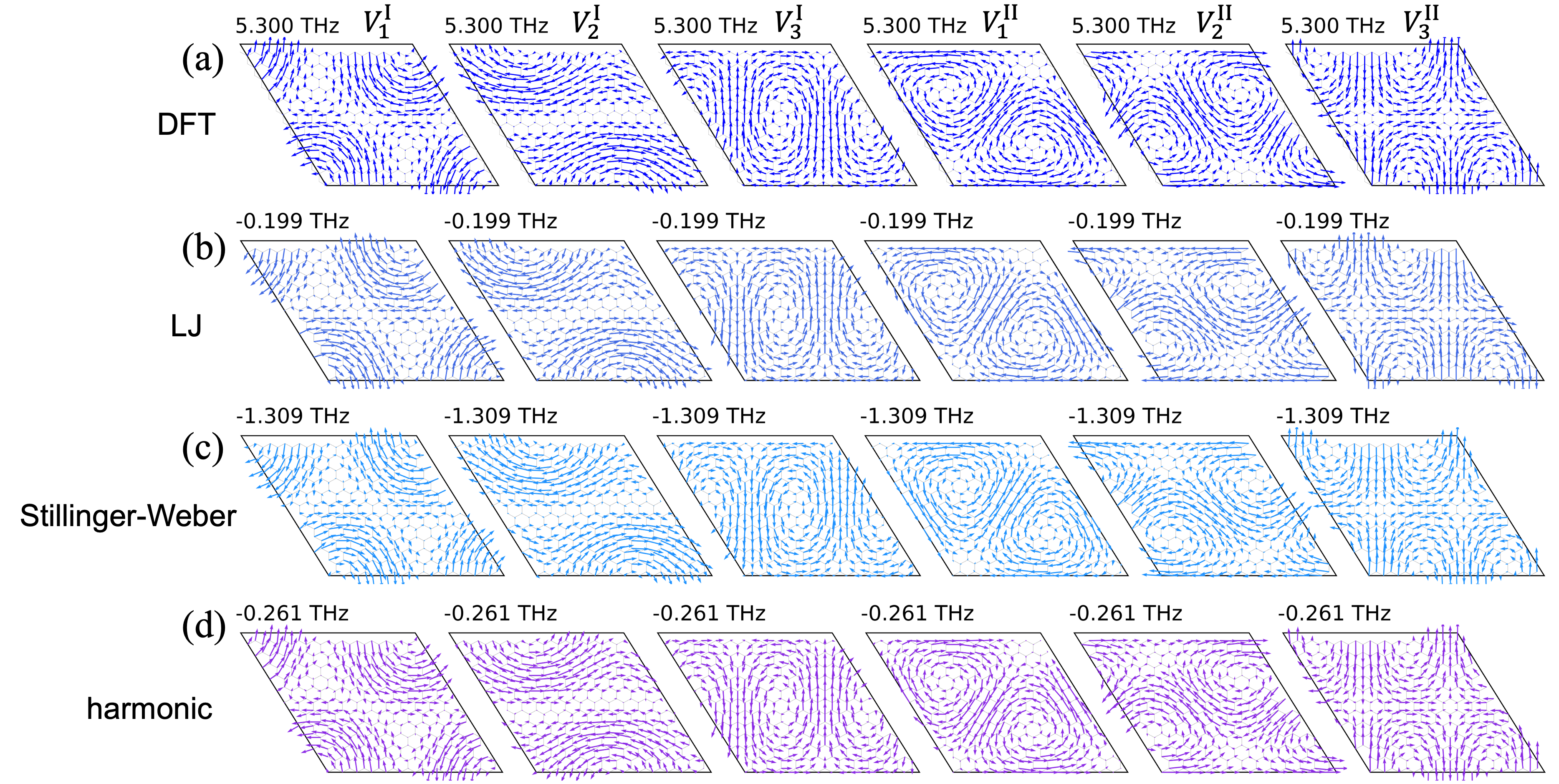}
\caption{The six lowest-frequency TA modes for a hexagonal lattice in a $12\times 12$ supercell. Results are obtained using (a) DFT, (b) the Lennard-Jones (LJ) potential, (c) the Stillinger--Weber potential, and (d) the harmonic potential (spring constant $k=1 \rm{eV/\mathring{\mathrm{A}}^2}$) with a cutoff $r_c=1.7 \mathring{\mathrm{A}}$ (the bond length is $1.4 \mathring{\mathrm{A}}$).}
\label{fig:S2}
\end{figure*}

We solve the phonon eigenvalue equation using various empirical potentials via phonoLAMMPS\cite{togo2023implementation,togo2023first}. For the hexagonal lattice, we employ the Lennard-Jones potential, a repulsive-only harmonic potential with a cutoff, and the Stillinger–Weber potential in LAMMPS\cite{thompson2022lammps,plimpton1995fast}. 
We also perform calculations at the quantum-mechanical level using density functional theory (DFT), as implemented in VASP \cite{kresse1996efficiency,perdew1996generalized}. The resulting vortex modes are plotted in Fig.~\ref{fig:S2}, demonstrating their independence from the choice of potential.


\clearpage

\section{High frequency phonon vortices}
\label{sec:SM3}

As a general principle, the TA/LA vortex modes at high frequencies are also formed by linear combinations of the planar modes, using the same SALC method as discussed in the main text.

\begin{figure}[htb]
\centering
\includegraphics[width=0.90\columnwidth]{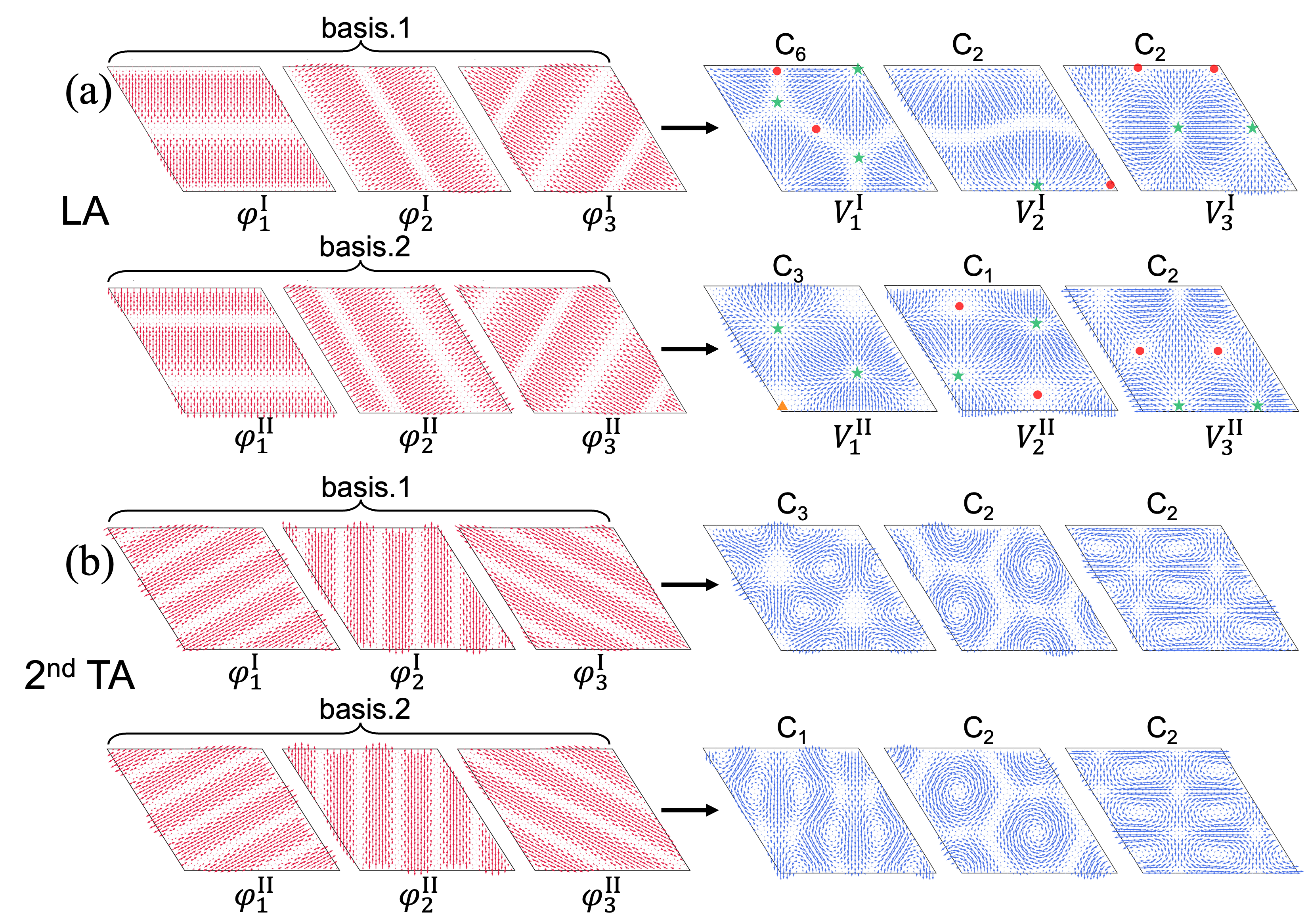}
\caption{High frequency phonon vortices. (a) the lowest LA modes at $\omega \approx 4.05$ THz and the corresponding planar modes. (b) the second lowest TA modes at $\omega \approx 4.07$ THz and the corresponding planar modes. 
}
\label{fig:S4}
\end{figure}

\clearpage

\section{Fixed boundary conditions}
\label{sec:SM4}
\enlargethispage{4\baselineskip}

\begin{figure}[tbh]
\centering
\includegraphics[width=0.75\columnwidth]{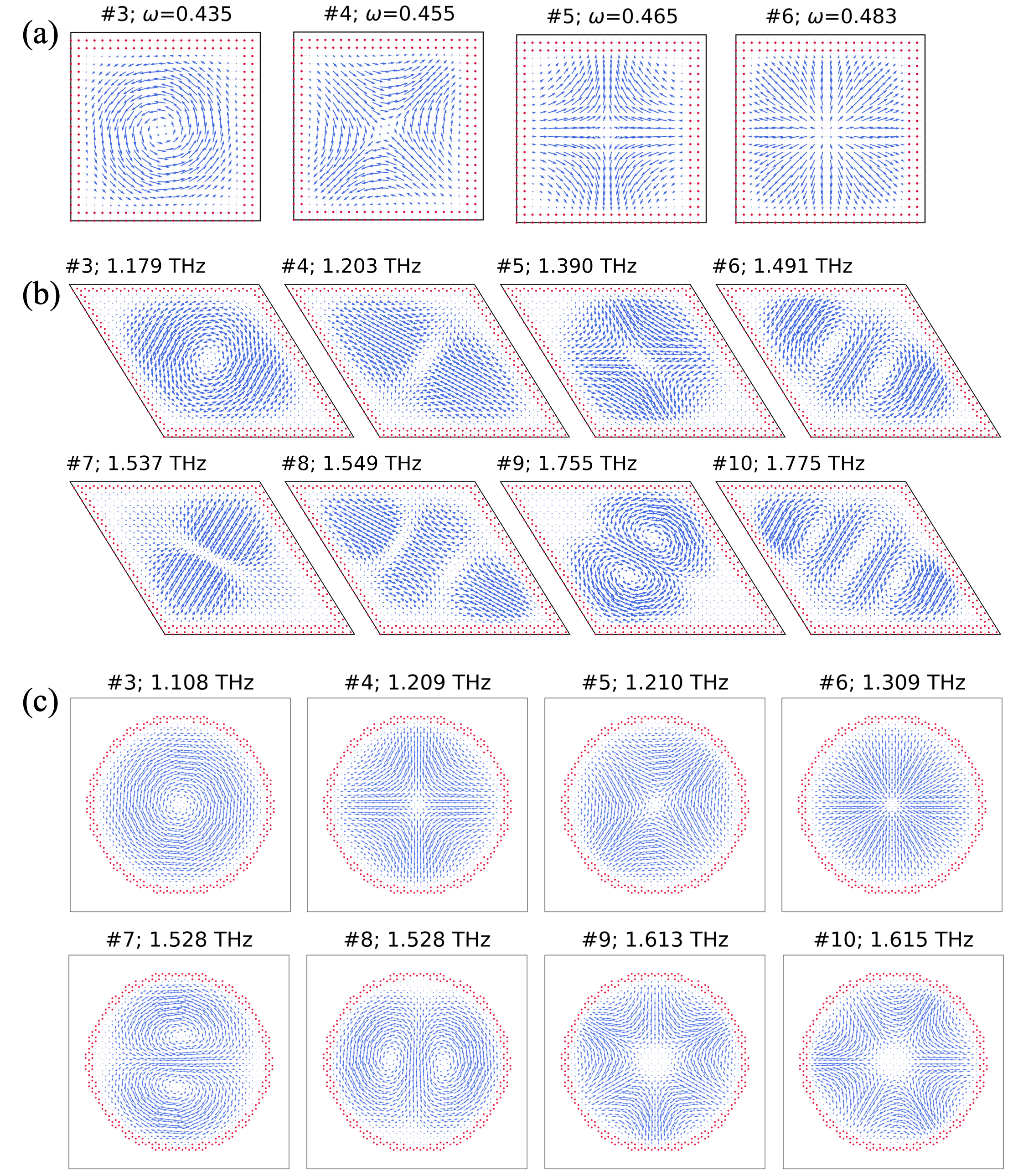}
\caption{Phonon vortices under fixed boundary conditions. We show eight lowest-frequency modes for 
(a) a $24 \times 24$ square lattice, (b) a $24 \times 24$ hexagonal lattice, and (c) a circular supercell of a graphene (Gr) lattice in a similar size of $R = 14 a$, where $R$ is the radius of the circular plate, and $s$ is the unitcell size. Red atoms near the boundaries are fixed.
}
\label{Fig.6}
\end{figure}

The phonon vortices with fixed boundary atoms are shown in Fig.~\ref{Fig.6}; their degeneracy is lifted, and the TA and LA modes become mixed. In general, the mode symmetry is now determined by the boundary conditions rather than by the lattice symmetry.

\FloatBarrier
\phantomsection
\label{sec:SMreferences}

\putbib[ref_SM]
\end{bibunit}

\end{document}